\documentclass[prd,twocolumn,superscriptaddress,nofootinbib]{revtex4-1}
\usepackage{epsfig}
\usepackage{graphicx}
\usepackage[english]{babel}
\usepackage{hyphenat}
\usepackage{amsmath}
\usepackage{amssymb}
\usepackage{mathtools}
\usepackage{mathrsfs}
\usepackage{slashed}
\usepackage{epstopdf}
\usepackage[dvipsnames]{xcolor}
\usepackage{booktabs}
\usepackage{subfigure}
\usepackage{float}

\usepackage{url}
\usepackage{breakurl}

\definecolor{lcolor}{rgb}{0.,0.0,0.}
\definecolor{citcolor}{rgb}{0,0.,0.5}
\usepackage[breaklinks,colorlinks,urlcolor=blue,citecolor=blue,linkcolor=blue]{hyperref}
\usepackage{multirow}
\usepackage{ltablex}
\usepackage{soul}
\usepackage{braket}
\usepackage{hepunits}

\colorlet{darkgreen}{green!50!black}
\colorlet{darkblue}{blue!70!black}
\colorlet{brightyellow}{yellow!75!red}
\colorlet{orange}{red!50!yellow}
\colorlet{darkgray}{gray!50!black}

\newcommand*\diff{\mathop{}\!\mathrm{d}}
 \newcommand{\GeV}{{{\,}\textrm{GeV}}}

\def\rrangle{\rangle\!\rangle}
\def\llangle{\langle\!\langle}
\newcommand{\beq}{\begin{eqnarray}}
\newcommand{\eeq}{\end{eqnarray}}

\newcommand{\bem}{\begin{multline}}
\newcommand{\eem}{\end{multline}}
\newcommand{\beg}{\begin{gather}}
\newcommand{\eeg}{\end{gather}}

\newcommand{\nn}{\nonumber\\}

\newcommand{\ben}{\begin{eqnarray*}}
\newcommand{\een}{\end{eqnarray*}}

\newcommand{\eqn}[1]{Eq.~\eqref{#1}}

\def\cW{{\cal W}}

\def\cT{{\cal T}}

\newcommand{\secn}[1]{Section~1}
\newcommand{\appn}[1]{Appendix~1}

\long\def\comment#1{ }

\def\and{\quad\text{and}\quad}

\def\q{{\boldsymbol q}}
\def\0{{\boldsymbol 0}}
\def\p{{\boldsymbol p}}
\def\c{{\boldsymbol c}}
\def\k{{\boldsymbol k}}

\def\n{{\boldsymbol n}}
\def\x{{\boldsymbol x}}
\def\y{{\boldsymbol y}}

\def\z{{\boldsymbol z}}

\begin{document}

\title{Medium induced jet broadening in a quantum computer}

\author{Jo\~ao Barata}
\email[]{jlourenco@bnl.gov}
\affiliation{Physics Department, Brookhaven National Laboratory, Upton, NY 11973, USA}
\author{Xiaojian Du}
\email[]{xjdu@physik.uni-bielefeld.de}
\affiliation{Fakult\"{a}t fur Physik, Universit\"{a}t Bielefeld, D-33615 Bielefeld, Germany}
 \author{Meijian Li}
\email[]{meijian.li@usc.es}
\affiliation{Instituto Galego de Fisica de Altas Enerxias (IGFAE), Universidade de Santiago de Compostela, E-15782 Galicia, Spain}
\author{Wenyang Qian}
\email[]{wqian@iastate.edu}
\affiliation{Department of Physics and Astronomy, Iowa State University, Ames, IA 50010, USA}
\affiliation{Instituto Galego de Fisica de Altas Enerxias (IGFAE), Universidade de Santiago de Compostela, E-15782 Galicia, Spain}
\author{Carlos A. Salgado}
\email[]{carlos.salgado@usc.es}
\affiliation{Instituto Galego de Fisica de Altas Enerxias (IGFAE), Universidade de Santiago de Compostela, E-15782 Galicia, Spain}

\begin{abstract}
QCD jets provide one of the best avenues to extract information about the quark-gluon plasma produced in the aftermath of ultra relativistic heavy ions collisions. The structure of jets is determined by multiparticle quantum interference hard to tackle using perturbative methods. When jets evolve in a QCD medium this interference pattern is modified, adding another layer of complexity. By taking advantage of the recent developments in quantum technologies, such effects might be better understood via direct quantum simulation of jet evolution. In this work, we introduce a precursor to such simulations. Based on the light-front Hamiltonian formalism, we construct a digital quantum circuit that tracks the evolution of a single hard probe in the presence of a stochastic color background. In terms of the jet quenching parameter $\hat q$, the results obtained using classical simulators of ideal quantum computers agree with known analytical results. 
With this study, we hope to provide a baseline for future in-medium jet physics studies using quantum computers.
\end{abstract}

\maketitle

\section{Introduction}
\label{sec:intro}
In recent years, there has been an increasing interest in applying novel developments in quantum information science to other scientific areas, in particular high energy physics (HEP), where many directions have been explored~\cite{Bauer:2022hpo}. 

Some of these novel proposals to use quantum technologies have ranged from the simulation of  scalar~\cite{Jordan:2012xnu,Jordan:2011ci,Brennen:2014iqu,Jordan:2017lea,Klco:2018zqz,Klco:2020aud,DeJong:2020riy,Barata:2020jtq,Yeter-Aydeniz:2021mol,Kurkcuoglu:2021dnw}, fermionic~\cite{Jordan:2014tma,Kharzeev:2020kgc,Bringewatt:2022zgq} and gauge field theories~\cite{Klco:2018kyo,Klco:2019evd,Chakraborty:2020uhf,Shaw:2020udc,Shaw:2020udc,Stetina:2020abi,Ciavarella:2021nmj,Honda:2021aum,Xu:2021tey,Ciavarella:2021lel,Nguyen:2021hyk,Rahman:2022rlg,Farrell:2022wyt,Atas:2022dqm,Nguyen:2021hyk} to thermal systems~\cite{Czajka:2021yll} and the  thermalization of non-equilibrium systems~\cite{Lamm:2018siq,deJong:2021wsd,Zhou:2021kdl}. Beside field theory based simulations, they have also been applied to specific topics such as nuclear structure~\cite{Mueller:2019qqj,Lamm:2019uyc,Kreshchuk:2020dla,Kreshchuk:2020aiq,Li:2021kcs,Qian:2021jxp}, neutrino oscillation~\cite{Arguelles:2019phs} and string theory~\cite{Gharibyan:2020bab}. Concerning collider oriented physics, 
these technologies have, for example, been used to simulate hard probes like heavy flavors~\cite{Gallimore:2022hai} and jets~\cite{Barata:2021yri,Yao:2022eqm}, optimize parton showers~\cite{Bauer:2019qxa, Williams:2021lvr,Gustafson:2022xwt} and jet clustering algorithms~\cite{Wei:2019rqy, Pires:2021fka,deLejarza:2022bwc}
as well as in the detection of quantum anomalies~\cite{Alvi:2022fkk} and the study of spin correlations at high energies~\cite{Gong:2021bcp}. 
Although such applications are still highly constrained by the performance of current quantum computers~\cite{Preskill_2018}, even the (re)formulation of problems in a language accessible to these machines turns out to be highly non-trivial. 
Seeing the expected melioration of quantum technologies in the next decades~\cite{ibm_road_map,honeywell_road_map}, the current conceptual work and small-scale implementations will prove crucial for the success of future large-scale applications.

One of the most important experimental programs being pursued in HEP is the ultra-relativistic collision of heavy ions~\cite{Busza:2018rrf}.
In the aftermath of such collisions, one can observe the production of the quark-gluon plasma, a state of matter expected to have existed in the first few microseconds of the universe. Experimentally, the properties of such plasmas can only be indirectly extracted by studying the yield and properties of a limited number of hard probes self-generated in each collision~\cite{Apolinario:2022vzg,Enterria2010}. One of the most successful and widely studied probes are QCD jets. Due to their multiparticle and multiscale structure, jets are by nature complicated objects to understand, even in a vacuum environment. When immersed in a dense background, their properties can be drastically modified and the success of the heavy ion physics program requires 
a clear understanding of such effects. The collection of medium induced jet modifications is colloquially referred to as jet quenching.

In the traditional picture of jet quenching, most jets' modifications result from the  interaction with a nearly thermalized background, which admits a classical description. On the other hand, jets are quantum objects, so studying them requires quantum field theory techniques. This dichotomy between the medium and jets' nature motivated a hybrid quantum strategy to study jet evolution in the medium~\cite{Barata:2021yri} (see also Ref.~\cite{DeJong:2020riy}). The focus was put on understanding the diffusion of a single parton (i.e. jet constituent) due to the multiple interactions with the background, but ignoring the production of induced radiation.

In this work, we implement these ideas to simulate medium-induced jet broadening on a quantum computer. Our formulation is based on a non-perturbative light-front Hamiltonian approach, the time-dependent basis light-front quantization (tBLFQ)~\cite{Zhao:2013cma, Li:2020uhl, Li:2021zaw}. 
We simulate the real time evolution of the jet in the medium at the amplitude level, extracting physical observables directly from the quantum state. 
A similar time-dependent quantum algorithm has been applied to solve the nuclear inelastic scatterings~\cite{Du:2020glq}.
Although the current problem is easily solved using analytical techniques, as detailed below, our quantum implementation provides a baseline for future simulations including, for example, multigluon production. Such higher order effects are already hard to completely tackle using traditional approaches, and it is there where quantum technologies might find room to prosper. We perform the quantum simulations using the IBM quantum framework {\tt qiskit}~\cite{Qiskit}. From the final quantum state, we extract the transverse momentum distribution and the associated jet quenching coefficient, $\hat q$, which plays a central role in phenomenology.

This manuscript is organized as follows. In Sect.~\ref{sec:theory}, we review the formulation of single parton evolution in the presence of a dense medium and we introduce the quantum simulation algorithm. Section~\ref{sec:to_qcomputer} provides a detailed description on how to formulate real time evolution of a single parton in the medium using a digital quantum computer. In Sect.~\ref{sec:results}, we present numerical results of this approach via available quantum simulators provided by {\tt qiskit}. A summary of the results and future avenues of research are discussed in Sect.~\ref{sec:conclusion}.

\section{Theoretical Setup}
\label{sec:theory}


In this section, we first review 
the theoretical formulation of a high-energy parton evolving through a medium, using a non-perturbative light-front Hamiltonian approach.
 The presentation closely follows our previous works~\cite{Li:2020uhl, Barata:2021yri}.
We then outline the framework of the quantum simulation algorithm~\cite{Feynman:1981tf,Georgescu:2013oza}, while leaving the details of our implementation to the next section.

\subsection{Parton evolution in the Hamiltonian formalism}\label{sec:theory_c}

\begin{figure}[h]
    \centering
    \includegraphics[width=0.5\textwidth]{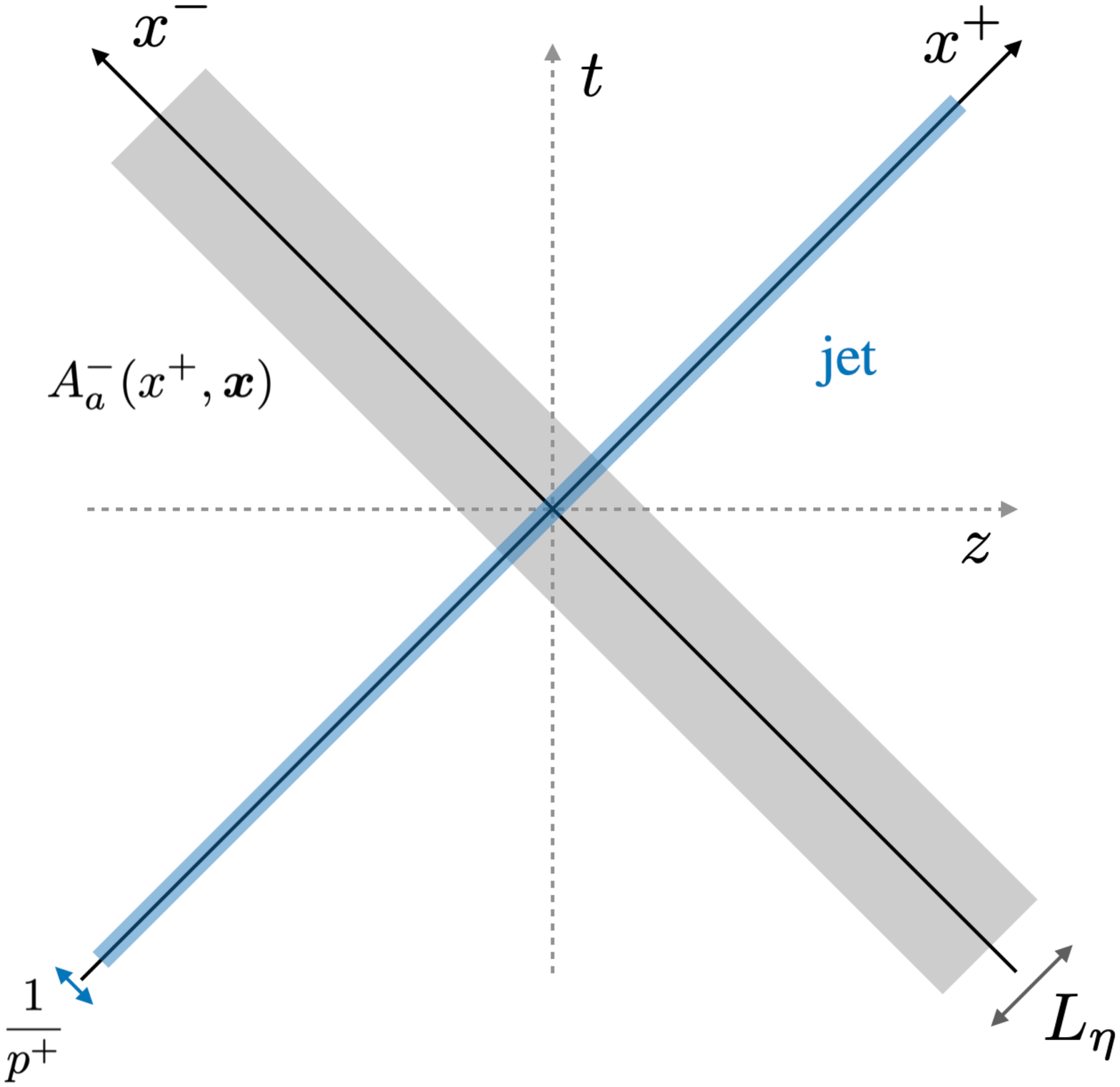}
    \caption{An illustration of the jet evolution in the presence of a highly boosted background plasma described by a classical field $A_a^\mu$.}
    \label{fig:event_picture}
\end{figure}

We consider the propagation of a highly energetic massless parton with light-cone momentum $p=(p^+,\p,p^-)$, moving close to the light-cone along the $x^+$ direction (see App.~\ref{appendix:conventions} for conventions of coordinates in this paper). The evolution of this hard probe occurs in the presence of a dense medium, which can be taken to be boosted along the $x^-$ direction. The medium is assumed to have a finite length $L_\eta$ along $x^+$. This process is illustrated in Fig.~\ref{fig:event_picture}. 

We treat the probe as a completely quantum object, whereas the medium, due to its large number of degrees of freedom, is described in this picture by a classical stochastic field $A^\mu(x)$. Notice that this approximation should be valid to characterize the quark-gluon plasma produced after high energy heavy ion collisions at RHIC or the LHC,\footnote{Though in this work we are interested in the evolution of jets in the quark-gluon plasma, the same formulation also applies to jet evolution in cold nuclear matter.} where jet quenching effects have been widely observed~\cite{CMS:2016uxf,STAR:2002svs}.

We consider the quark jet in the leading Fock sector, and the Hamiltonian (denoted as $\hat P^-$) can be obtained following the canonical light-front quantization formalism~\cite{Brodsky:1997de,Blaizot:2015lma,Li:2020uhl}. 
It can be split into two different terms
\begin{align}
\hat P^-\equiv  \hat K + \hat V\, .
\end{align} 
In the absence of the background medium, the real time evolution of the jet is solely controlled by the kinetic energy operator~\cite{Brodsky:1997de,Blaizot:2015lma}
\begin{align}\label{eq:K}
\hat K=  \frac{\hat \p^2}{2p^+} \; ,
\end{align}
corresponding to the spatial quantum diffusion 
of the probe at a fixed light-cone energy $p^+$.\footnote{We write the Hamiltonian in the operator form, and its full expression, in terms of field operators, can be found in Ref.~\cite{Li:2020uhl}.
} In the presence of a medium, the interaction term also comes into play,
\begin{align}\label{eq:Vdef}
\hat V= g \hat A^-_a T^a\;,
\end{align}
with $T^a$ denoting the color generators in the probe's color representation. 
At high energies, where $p^+\gg p_\perp>p^-$, the probe is highly localized around $x^-=0$ and one can simplify the field's spacetime dependence to be $A^\mu(x) \approx A^\mu(x^+,\x) $.
Additionally, the structure describing local parton-medium interactions, $\overline \Psi(p-q) \gamma^\mu A_\mu^a(q) T^a \Psi(p)$, with $q$ the exchanged momentum with the medium, only receives contributions from the $\mu=+$ component~\cite{Brodsky:1997de,Mehtar-Tani:2013pia,Casalderrey-Solana:2007knd,Kovchegov:2012mbw}, up to power suppressed terms in the jet energy $p^+$. 
Therefore, the probe evolution in the medium is only sensitive to the component $A^-$, as in Eq.~\eqref{eq:Vdef}. 

In modeling the statistics of the stochastic background field, we take the simplest and most widely used approximation, which assumes that the plasma's color charges are completely uncorrelated and have white-noise statistics. This corresponds to an extended version of the McLerran-Venugopalan (MV) model~\cite{McLerran:1993ka,McLerran:1993ni}, where the field satisfies the reduced classical Yang-Mills equation,\footnote{We consider the MV model since it is widely used and it also allows for a simple comparison to analytical results. Nonetheless, our quantum implementation is more general and others models could be considered.} 
\begin{align}\label{eq:poisson}
 (m_g^2-\nabla^2_\perp )  A^-_a(x^+,\x)=\rho_a(x^+,\x)\;,
\end{align}
with $m_g$ an effective gluon mass introduced to regularize the infrared (IR) divergence of the field and ensuring the color neutrality of the source distribution~\cite{krasnitz2003gluon}. The charge density $\rho_a$ describes the energetic degrees of freedom generating the field, and its only non-trivial correlator reads
\begin{multline}\label{eq:chgcor}
 \llangle \rho_a(x^+,\x)\rho_b(y^+,\y) \rrangle \\=g^2 \mu^2(\x)\delta_{ab}\,\delta^{(2)}(\x-\y)\,\delta(x^+-y^+)\;.
\end{multline}
Here and throughout this paper, we use $\llangle \cdots \rrangle$ to denote the average over medium configurations. The quantity $\mu$ can be understood as the density of scattering centers in the medium, which dictates the strength of the parton-medium interaction. 
The resulting interaction between the medium and the probe is local in position space.

The time evolution of the probe, in the Schr\"{o}dinger picture, is controlled by the time evolution operator $U$,  such that the 
\begin{align}~\label{eq:master_eq}
\begin{split}
 \ket{\psi_{L_\eta}}=& U(L_\eta;0)\ket{\psi_0}\\
 \equiv& \cT_+ e^{-i\int_{0}^{{L_\eta}} \diff x^+\, P^-(x^+) } \ket{\psi_0} \;,
\end{split}
\end{align}
where $\cT_{+}$ is the light-front time ordering operator, and $\ket{\psi_{x^+}}$ the quantum state of the jet at time $x^+$.  
From here, one can see that the problem we are solving is equivalent to a $2+1$ dimensional quantum mechanical problem, with the associated Hilbert space being that of a single particle in a two-dimensional (transverse) space. We solve~\eqn{eq:master_eq} through a non-perturbative treatment, decomposing the time-evolution operator as a sequence of small steps in the light-front time $x^+$
\begin{align}~\label{eq:U_tBLFQ}
\begin{split}
 U(L_\eta;0)
=& \prod_{k=1}^{N_t} U(x^+_k;x^+_{k-1})\;,
\end{split}
\end{align}
with the intermediate times defined as $x^+_k=k\, L_\eta/N_t$ 
, and $N_t$ the total number of time segments.
In this way, the smallness of the step size allows to approximate the exact evolution operator by a product formula 
within each step, with the full evolution in whole being non-perturbative.
By doing so, we simulate the evolution of a quantum state at amplitude level. 
We will address the details of the implementation of the quantum circuit in the next section.

With the quantum state obtained from the simulation, one can directly study any given observable $\hat O$ from its expectation value $\langle \hat{O}(x^+) \rangle\equiv\braket{\psi_{x^+}|\hat O|\psi_{x^+}}$.
For any observable $\hat O$, the field in a single simulation is generated from a stochastic source configuration, i.e. $\rho_a$ satisfying Eq.~\eqref{eq:chgcor}, and one should take the configuration average at the observable level,
\begin{align}
    \llangle \langle \hat {O} \rangle  \rrangle =\lim_{n\to \infty}\frac{1}{n}\sum_{i=1}^n\langle \hat {O} \rangle _i\; ,
\end{align}
where $\langle \hat{O} \rangle_i$ denotes the expectation value for the $i^{\rm th}$ field configuration.
In reality, one can think of running the simulation multiple times with different field configurations as having multiple events in an actual collision experiment, with field fluctuations corresponding to the randomness of the medium in each event.

 In this work we will be interested in the case where $\hat O= \hat \p^2$. The medium averaged expectation value of this observable can be related to the jet quenching parameter via
\begin{align}\label{eq:def_qhat}
\hat q \equiv \frac{\llangle \langle \hat \p^2 (L_\eta)  \rangle \rrangle-\llangle \langle \hat \p^2 (0)  \rangle \rrangle }{L_\eta}  \, ,
\end{align}
corresponding to the average squared transverse momentum acquired per unit length.\footnote{The definition of $\hat q$ is not unique and dependent on the particular regularization used for the underlying ultraviolet diverge. We refer to reader to Refs.~\cite{Majumder:2012sh,Liu:2006ug,Benzke:2012sz,Casalderrey-Solana:2007ahi} for a more detailed discussion on the definition of $\hat q$.} This coefficient is responsible for describing both the diffusion of particles in the medium and radiative energy loss~\cite{Mehtar-Tani:2013pia}.

\subsection{Quantum simulation algorithm}\label{sec:theory_q}

The quantum simulation algorithm, as pioneered by Feynman~\cite{Feynman:1981tf}, allows access to the real-time dynamics of a target quantum system by simulating them in another controllable quantum system. Although classical analogous of such an approach can be implemented, they entail a linear growth in hardware and simulation time with the system size. In contrast, using a quantum computer, the computational resources and simulation time are expected to only scale logarithmically. This so-called ``quantum speed up" stems from the possibility of being able to explore the quantum nature of such devices, although its realization is problem-dependent.

For the current problem of a bare quark evolving in a dense medium, using the quantum simulation algorithm is less likely to bring a significant computational enhancement compared to a classical simulation, especially when the problem size is relatively small. However, to extend the current picture to allow for gluon emissions, which we will address in the forthcoming work, the respective quantum algorithm will enable an exponential speedup compared to the classical counterpart.

Following Sect.~\ref{sec:theory_c}, the evolution of the probe in the medium can be mapped to a time-dependent quantum mechanical problem.
Thus, in principle, it can be solved using a digital quantum computer~\cite{nielsen_chuang_2010}, which can be ideally thought as a collection of $1/2$-spins (qubits) that interact with each other via a system Hamiltonian.
Such devices can be described using the quantum circuit formalism, where  the interactions between different qubits can be decomposed in terms of basic unitary operations named quantum gates.
The digital quantum simulation algorithm can be summarized in five generic steps (see also Chapter 4 of Ref.~\cite{nielsen_chuang_2010}):

\begin{enumerate}
    \item \textbf{Input}: description of the target quantum system in terms of the system Hamiltonian and underlying Hilbert space.  
    If the original system lives in a infinitely large Hilbert space, an adequate discretized version must be provided.\footnote{ In our case, the inputs regard the Hamiltonian, which we have detailed in Sect.~\ref{sec:theory_c}.}
    
    \item \textbf{Encoding}: mapping the degrees of freedom of the problem to qubits on the quantum computer.

    \item \textbf{Initial state preparation}:   preparing the computational initial state $\ket{\psi_0}$ from a fiducial state native to the computer.
    \item \textbf{Time evolution}: building the series of quantum gates representing the evolution operator $U(L_\eta;0)$.

    \item \textbf{Measurement}: extracting information from the final quantum state $\ket{\psi_{L_\eta}}$. 
\end{enumerate}

\section{Qubitization of in-medium jet evolution}
\label{sec:to_qcomputer}
Following the procedure outlined in the preceding section, Sect.~\ref{sec:theory}, we now implement the quantum algorithm for the problem of interest, the evolution of a quantum probe through the medium. We address four key elements of the implementation in this section.

 \subsection{Basis encoding}\label{sec:basis_encoding}
  We start by formulating the problem of in-medium single parton evolution in a lattice, such that the underlying Hilbert space becomes finite and can be mapped to the qubits in the quantum computer. We first build up the basis for the spatial part and then extend it by adding a color phase space. In this paper, we consider that the probe lives in either the U(1) or SU(2) fundamental color representations.\footnote{For QCD the relevant color group is SU(3). The SU(2) and SU(3) groups are both non-Abelian, and for the problem of interest we show that the results between the two groups only differ by a global Casimir color factor. Nonetheless, and as detailed in Ref.~\cite{Barata:2021yri}, the implementation of the quantum algorithm for the SU(3) case is technically more complicated.} For the U(1) case, it is sufficient to consider the spatial phase space. Choosing an encoding scheme is in the same spirit as selecting a basis representation in the classical counterpart using the tBLFQ approach~\cite{Zhao:2013cma, Li:2020uhl}.

In order to discretize the dynamics of the probe we introduce a two-dimensional transverse lattice, with a span of $2 L_\perp$ and a number of $2 N_\perp$ sites per dimension  such that the lattice spacing is $a_\perp=L_\perp/N_\perp$. We impose  periodic boundary conditions to the lattice. 
Any position state vector $\ket{\x}=\ket{x_1,x_2}$ describing the location of the probe can be mapped to a lattice vector $\ket{\n}=\ket{n_1,n_2}$, such that $\x= \n \, a_\perp$.
The reciprocal lattice corresponding to the transverse momentum lattice has spacing $b_\perp= 2\pi/ (2L_\perp)$. Similarly to the position space lattice, we can map a momentum state vector $\ket{\k}$ to a momentum space lattice vector $\ket{\q}=\ket{q_1,q_2}$, such that $\k=\q \, b_\perp$. The two reciprocal basis vectors are related through an inverse Fourier transformation (see also App.~\ref{appendix:conventions}),
    \begin{align}\label{eq:FT}
    \ket{\n}= \frac{1}{2N_\perp}\sum_\q e^{i 2\pi \q\cdot \n/(2 N_\perp)} \ket{\q} \, .
    \end{align}
It follows that for any quantum state $\ket{\psi}$ describing the jet, there is a corresponding discretized version on the lattice which can be written in terms of a finite superposition of lattice state vectors $\{\ket{\n}\}$ (or equivalently $\{\ket{\q}\}$).

Due to the periodic boundary conditions imposed on the lattice, two grid points which differ by an integer number of periods must be identified. As a result, for any operator, we have  
\begin{align}\label{eq:prdbd}
        \hat O \ket{ n_1, n_2}=\hat O \ket{ n_1 + i\,(2N_\perp), n_2+ j\,(2N_\perp)} \,,  
\end{align}
with $i,j=0,\pm 1, \pm 2,\ldots$, and likewise on the momentum lattice. To extract the physical information from the prepared quantum states, it is convenient to use  the fundamental Brillouin zone where $n_i, q_i \in \{-N_\perp, -N_\perp+1,\ldots, N_\perp-1\}$, with $i=1,2$.
On the other hand, when performing the quantum simulations, it is more advantageous to go to the Brillouin zone where $n_i, q_i \in \{0,1,\ldots, 2N_\perp-1\}$, with $i=1,2$. The advantage of this choice is two-folded. First, using non-negative integers, we can directly match each number in its decimal form to a non-negative binary representation. Each binary digit can be thought of as a spin-up state ($\ket{0}$) or spin-down state ($\ket{1}$). 
As an example, we can represent the state $\ket{1,3}$ (for either $\ket{\n}$ or $\ket{\q}$ basis) as
\begin{align}\label{eq:ooo}
\ket{1,3}\to  \ket{01,11}  &\to \ket{0}\otimes \ket{1}\otimes \ket{1}\otimes \ket{1} \nn         &\to  \ket{\uparrow}\otimes \ket{\downarrow}\otimes \ket{\downarrow}\otimes \ket{\downarrow}
 \; ,  
\end{align}
using four spins (qubits). 
Second, having both $\ket{\n}$ and $\ket{\q}$ on the fundamental Brillouin zone, 
we can convert between the two according to~\eqn{eq:FT} using the standard quantum Fourier transform (qFT) algorithm~\cite{nielsen_chuang_2010}. Using other Brillouin zones would require a modified version of the standard qFT algorithm~\cite{Klco:2018zqz}, incurring in extra quantum gate operations. 
Therefore, we formulate the simulation algorithm in the non-negative Brillouin zone, and interpret physical observables on the fundamental zone using the periodicity of the lattice as in Eq.~\eqref{eq:prdbd}.

Having encoded the transverse state $\{\ket{\n/\p}\}$ to qubits, we complete the basis encoding for the U(1) system. For SU(2), we need to further encode the color sector of the jet state. Since in the fundamental representation of SU(2) there are only two color states, the color sector can be described by a single qubit, where each classical spin state corresponds to a different color state. As such, the encoded SU(2) basis is $\{\ket{\n/\p}\}\otimes \ket{c}$ with $c=0,1$ denoting the color state. For the encoding of the color sector with a general SU($N_c$) group, we refer to Ref.~\cite{Barata:2021yri}.

The number of qubits required to encode the states per transverse dimension is determined by
\begin{align}\label{eq:nQ}
n_Q=\log_2 2N_\perp\, .
\end{align}
Therefore, we need a total of $2n_Q$ qubits to account for the two-dimensional lattice as in the U(1) case, and $2n_Q+1$ in the SU(2) case, with the extra qubit tracking the color state.

\subsection{Gate encoding and time evolution}\label{sec:gate encoding}

 Having discussed the discretization of the system in terms of qubits, let us now focus on the construction of the time evolution operator $U$ in terms of quantum gates.
 
 To this purpose, we implement the simplest product formula decomposition, splitting the evolution into $N_t\equiv L_\eta/\delta x^+$ steps of duration $\delta x^+$; see \eqn{eq:U_tBLFQ}. 
 For a truly evolving background, the Hamiltonian $\hat P^-$ can be time dependent if, for example, the medium becomes more dilute with time. In this work, we do not consider such scenarios and assume that the medium profile is constant in $x^+$. However, since we are dealing with a stochastic background, there is an emergent $x^+$ time dependence in the Hamiltonian. To numerically simulate this feature, we slice the medium into $N_\eta$ layers along $x^+$ \cite{Lappi:2007ku,Ipp:2020mjc}, such that the time duration for each layer is $\tau\equiv L_\eta/N_\eta$. The medium at different time layers is generated from independently sampled sources, thus ensuring that the correlators in \eqn{eq:chgcor} are satisfied 
 within a resolution window of $\tau$ in the $x^+$ dimension.
 
Within each small step $\delta x^+$, the Hamiltonian can be approximated as being constant in $x^+$. Implementing a first order Trotter decomposition in each step, the time evolution for each discretized time step, $U(x^+_{k+1};x^+_{k})$, is approximated as~\cite{Barata:2021yri}
\begin{align}\label{eq:U_deltat}
&U(x^+_k+\delta x^+;x^+_{k})\approx U_K(\delta x^+) \, U_A(\delta x^+,x_k^+) \nn 
&\equiv \exp\left\{-i \delta x^+ \frac{\hat \p^2}{2p^+}\right\}     \exp\left\{-ig \delta x^+ \hat A_a^-(x^+_{k}) T^a\right\} \, ,
\end{align}
with $k=1,2,\ldots, N_t$. 
Here, we denote the evolution operator according to the kinetic energy and the medium interaction, for a small time duration $\delta x^+$, as $U_K(\delta x^+)$ and $U_A(\delta x^+,x_k^+) $, respectively. 
This formula gets corrections $\mathcal{O}((\delta x^+)^2)$, and the full evolution $U(L_\eta;0)$ as a product is exact in the limit $N_t\to\infty$. 
Note that with this treatment, the duration of each time step $\delta x^+$ cannot be larger than $\tau$.

Noting that the $\hat \p^2$ and $\hat A_a^-$ operators are separately diagonal in the transverse momentum and position spaces, we implement~\eqn{eq:U_deltat} by first applying $U_K(\delta x^+)$ in the momentum basis $\{\ket{\q}\}$, and then $U_A(\delta x^+,x_k^+)$ in the coordinate basis $\{\ket{\n}\}$, using a Fourier transform in order to change basis $\{\ket{\q}\} \to \{\ket{\n}\}$. After $U_A$ acts on the state, we perform the transformation $\{\ket{\n}\} \to \{\ket{\q}\}$, and iterate the algorithm.  
In what follows, we detail the quantum gate encoding for the two parts of the evolution accordingly.

A straightforward way of implementing $U_K(\delta x^+)$ is to first decompose 
the $\hat \p^2$ operator as a sum of strings of Pauli operators, and then time-evolve the Paulis, e.g. with the {\tt PauliEvolutionGate} class in {\tt qiskit}, or making a use of further product formula decompositions~\cite{nielsen_chuang_2010}. 
However, as the dimension of the underlying lattice grows, such a decomposition becomes in general inefficient, since the number of strings and their powers increase linearly with the system size. Then, a potentially more efficient implementation consists in using a variation of the phase kickback algorithm~\cite{Barata:2021yri,nielsen_chuang_2010,Zalka:1996st,Wiesner:1996xg}. 
Such an approach could in principle shorten the circuit depth,  however, it would require using arithmetic gates to obtain the value of $\q^2$ from the $\q$-encoded qubit and extra ancilla qubits.

In this work, we perform the quantum simulations mainly using {\tt qiskit}'s classical backends for ideal quantum computers, requiring a small number of qubits.
Therefore, it is more favorable to use the direct exponentiation approach, since it requires no extra gates or qubits, unlike the above mentioned strategies. Also, the circuit depth using this approach is comparable to the other two strategies for the lattices ($N_\perp=16,32$) being considered. The other approaches only become competitive for larger size problems.

Our implementation of $ U_A(\delta x^+,x_k^+)$ takes the classical values of $A_a^-(x^+_{k}, \x)$ as an input~\cite{Barata:2021yri}.
As introduced in Sect.~\ref{sec:theory_c}, these amplitudes carry configuration-wise fluctuations,
and are generated beforehand in a classical computer following the procedure formulated in Ref.~\cite{Li:2020uhl, Li:2021zaw} (see also App.~\ref{app:field}).
This procedure introduces a classical cost when compiling the full algorithm.
However, for a wide range of parameters, this classical output can be generated in an economical fashion using standard personal computers. 
Provided the values of the medium, the simulation of $U_A(\delta x^+; x^+)$ itself is purely quantum-mechanical.
Let us now detail its implementation for the U(1) and the SU(2) gauge groups.

In the case of a U(1) medium, the evolution operator is diagonal in the full basis space $\{\ket{\n}\}$. Thus, the operator $U_A(\delta x^+; x^+)$ can be implemented using the phase kick-back approach discussed above in the context of $U_K$, with the same shortcomings (extra ancilla qubits and auxiliary operations).
An alternative to this method is to write a generic diagonal operator in terms of smaller gates, where the field values would be direct inputs to the algorithm. However, such an approach would require the application of a number of gates scaling linearly with the system size~\cite{bullock2003smaller,Welch_2014}. A more feasible and efficient approach, generalizing the ideas of~Refs.\cite{Zalka:1996st,Wiesner:1996xg}, consists in further discretizing the field values. With such an extra discretization step, it has been shown~\cite{Kassal_2008} that a diagonal operator can be implemented efficiently at a cost of extra ancilla qubits and applications of qFTs.  
In carrying the quantum simulations, we take a small number of qubits, and we find it most efficient to implement $U_A$ for the U(1) case by direct exponentiation of field value matrix. In {\tt qiskit}, the equivalent gate is constructed following the quantum Shannon Decomposition algorithm~\cite{Shende2006} using the native {\tt Operator} class. As a result, when transpiling $U_A$ to a real device, the length of the circuit becomes much longer than the maximal coherence time of any available quantum computer.

Following the gate encoding of $U_A(\delta x^+; x^+)$ in the U(1) case, we make the implementation for SU(2) by adding the color sector. The respective operator is $2\times 2$ block-diagonal in the color transverse coordinate basis, as formulated in Sect.~\ref{sec:basis_encoding}.

We implement $U_A$ in the SU(2) scenario by also using direct exponentiation, which results in an exact implementation in each time step.
Similar to the previous operator, one could implement similar strategies to the ones discussed above for a more efficient approximate implementation of the evolution operator. Here we want to highlight a particularly interesting modular approach, which splits position/momentum and color space, allowing to reuse the U(1) implementation. This is achieved by doing a product decomposition of $U_A$, such that each term is controlled by a single color matrix. Then the evolution in position space is captured the exact U(1) evolution matrix, with a control based on the color qubit; see Ref.~\cite{Barata:2021yri} for a more detailed discussion.  
Despite the theoretical advantages of having modularized quantum circuits and reduced number of qubits in the evolution, this selection operation in practice can be expensive by introducing a large number of single qubit controls in the circuit. We found that the overall simulation time actually increases dramatically for a reasonable lattice size of $N_\perp=16$ compared to the exact implementation. Therefore,  all the simulation results we present in this work use direct matrix exponentiation using the native {\tt Operator} class, which is optimal given the scale of our problem.

  \subsection{Initial state preparation}

  Using the basis encoding detailed in Sect.~\ref{sec:basis_encoding}, one can, in principle, prepare any initial state $\ket{\psi_0}$ in terms of superposition of basis states, though the preparation of an arbitrary initial state might not always be achieved efficiently in practice.
 Since we are interested in studying the jet evolution in momentum space neglecting initial state effects, we take $\ket{\psi_0}$ to be the zero transverse momentum state, i.e. $\ket{\q=\mathbf{0}}$. 
 At the quantum circuit level, such a state corresponds to all qubits being in the $\ket 0$ state. 
 Then, for the color sector in the SU(2) case, we act on the color qubit with a Hadamard gate~\cite{nielsen_chuang_2010}, thereby generating a fully balanced superposition color state. It should be noted that a Gaussian initial state is also a common choice to study jet broadening~\cite{Sadofyev:2021ohn}, and can be prepared using known quantum algorithms~\cite{KitaevGauss,Deliyannis:2021che}.

\subsection{Measurement}\label{sec:method_measure}

The last key element in our quantum algorithm is the measurement, where we extract the information about the transverse momentum distribution from the final quantum state  $\ket{\psi_{L_\eta}}$. 
At the end of each quantum simulation, the prepared quantum state is measured, collapsing to a momentum eigenstate. By performing multiple measurements (shots), we are able to reconstruct the distribution of the jet state in momentum space.
The required number of shots would grow linearly with the system size if one maintains the desired measurement accuracy. Similarly, this projective measurement approach would become more resource-demanding if one wants to make a higher precision measurement.
Having reconstructed the underlying momentum distribution, its expectation values can be computed classically, from which the quenching parameter $\hat q$, as defined in \eqn{eq:def_qhat}, can be readily extracted.

Besides this approach to measuring the quantum state, there are more efficient strategies to extract expectation values of local operators without having to construct the classical probability distribution function. 
However, these strategies would, in general, lead to an increase in the circuit length or require the usage of extra ancilla qubits. Thus for the small circuits we construct, it is preferable just to perform a large number of shots.
Nevertheless, certain strategies with their unique features are still worth noting and could potentially become preferable once more powerful quantum devices are available in the future. In the following, we lay out three further strategies to measure $\hat q$.

The first strategy uses the so-called Hadamard test to access the real and imaginary parts of operator expectations value, and has been discussed in Ref.~\cite{Barata:2021yri}. The second alternative would be to consider the expectation value $\braket{\psi_{L_\eta}|\hat \p ^2|\psi_{L_\eta}}$, used in the definition of $\hat q$ given in \eqn{eq:def_qhat}. 
One could, for example, first prepare two states: the final state $\ket{\psi_{L_\eta}}=U(L_\eta; 0)\ket{\psi_0}$ and the state $\hat \p ^2\ket{\psi_{L_\eta}}$ in the quantum computer. Then, using a SWAP test (see Ref.~\cite{Foulds:2020ajt} and references therein), the contraction of these two states gives access to $\hat q$. However, $\hat \p$ is not an unitary operator and its naive implementation in terms of quantum gates is not possible. To solve this issue, one can either rewrite $\hat \p^2$ as a sum of Pauli strings or use more sophisticated techniques that allow to implement Hermitian operators under certain conditions~\cite{Robin2014,Berry_2015}. Third and last, another possible strategy that may be used to extract $\hat q$
takes advantage of the discretized form of the correlator in \eqn{eq:qhat_def_cont} with classically-computed Fourier transforms; see Ref.~\cite{Lamm:2019uyc} for further discussion.

As aforementioned, an efficient implementation of those alternative strategies is non-trivial and would require extra quantum resources or running time.  
With our method of measuring $\hat q$, an optimal choice at the current stage, we can proceed and extract the necessary information from the final state.

\section{ Quantum simulations and results}
\label{sec:results}

In this section, we study the quantum simulation results for the evolution of a quark jet in a dense stochastic medium, using the method formulated in the preceding sections.
The simulation accounts for both the quark quantum diffusion and soft gluon interactions with the medium as a static background field. We assume the medium to be homogeneous (more precisely, both $g^2\mu$ and $m_g$ spatially constant) in the transverse plane, except in Sect.~\ref{sec:res_anisotropic} when we study the effect from an anisotropic medium~\cite{Sadofyev:2021ohn,Barata:2022krd, Fu:2022idl}.
We perform the simulations using ideal QASM simulators from \texttt{qiskit} on both U(1) and SU(2) media. With the simulation results, we study the effect of quark jet momentum broadening, in particular the quenching parameter $\hat q$ at different saturation scales $Q_s^2$, as discussed in Sects.~\ref{sec:res_U1}, \ref{sec:res_SU2}.
Lastly, in Sect.~\ref{sec:res_noise}, we examine the results from simulations with quantum noise. For this purpose, we run the circuit using QASM simulators with an underlying noise model and using a real quantum IBM processor.

In the simulations, we take $L_{\perp}=4.8~\GeV^{-1}$ such that the medium extends transversely about 2 fm. We fix the strong coupling to $g=1$. The duration of the medium is taken to be $L_{\eta}=50~\GeV^{-1}$= 9.87 fm, with the number of layers $N_\eta=64$, and the IR regulator $m_g=0.8~\GeV$, except if mentioned otherwise.
We take $N_\perp=16$, and the number of qubits required in these simulations is therefore $ 2n_Q=
10 $ for the U(1) and 11 for the SU(2) case, according to Eq.~\eqref{eq:nQ}. 
For the charge density $g^2\mu$, which characterizes the strength of the medium, we take four representative values such that the saturation scale $Q_s^2$ defined as 
\begin{align}
Q_s^2\equiv C_F \frac{g^4 \mu^2 L_{\eta}}{2\pi } \, ,
\end{align}
 is in the range of $ 5-35 \GeV^2$. Here we introduce the fundamental Casimir $C_F\equiv (N_c^2-1)/{(2N_c)}$ for an SU($N_c$) group, such that $C_F=3/4$ for SU(2), and we take $C_F=1$ for a U(1) gauge group . 
It should be noted that since the transverse basis is discrete and finite, a pair of IR and UV cutoffs exist naturally (see also Sec.~\ref{sec:basis_encoding}). 
In determining the values of the lattice spacing $a_\perp$, we have taken into account that lattice effects are mitigated and the relevant physics is captured; see App.~\ref{app:criteria} for details.

Before presenting the simulation results, let us examine the uncertainties associated to the output measurement. In measuring an observable, an essential difference between the quantum simulation and its classical counterpart is that the former makes measurements on a quantum state, whereas the latter operates on its obtained projected wavefunction. Consequently, a large number of shots needs to be taken to extract a single observable, as discussed in Sect.~\ref{sec:method_measure}. 
This statistical measurement of the quantum circuit output is analogous to experimental measurements of jets over multiple events.
Accordingly, we quantify the uncertainties from our quantum measurements as statistic uncertainty. It should be noted that this uncertainty is different from the configuration fluctuations that arise from the stochastic medium. 

\begin{figure}[H]
    \centering
    \includegraphics[width=0.43\textwidth]{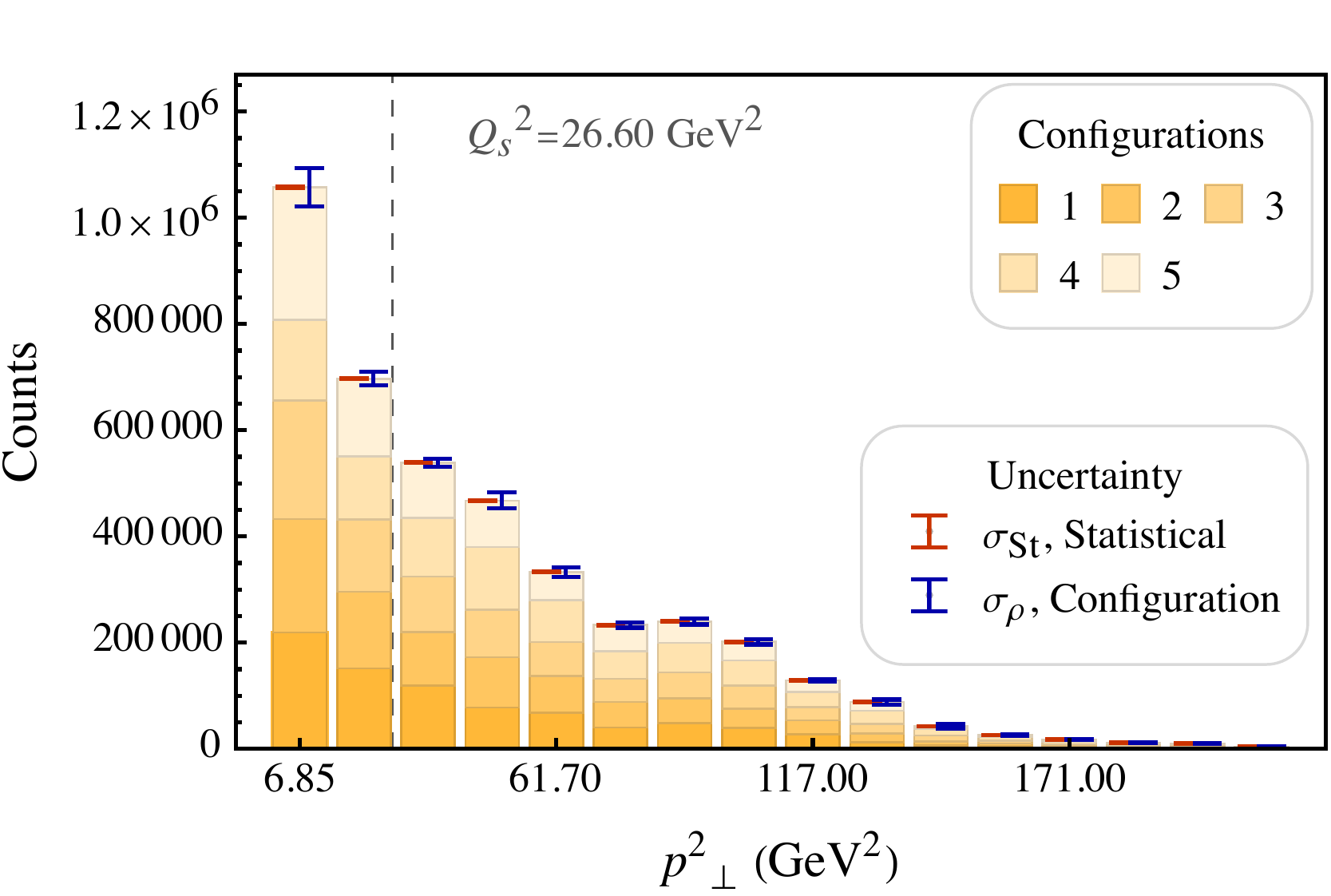}
    \caption{
    Number of counts (shots) taken while measuring $\hat \p^2$ on the quantum circuit for five different medium configurations, stacked in each bin. The dashed vertical line indicates the value of $Q_s^2$.
    The statistical and configuration uncertainties are indicated by red and blue error bars, respectively. See main text for definitions and details.
     }
    \label{fig:error_stat_config}
\end{figure}

As a concrete example, we present a measurement of $\hat \p^2$ by taking 819200 shots for each of five different medium configurations, using $N_\perp=32$ and $Q_s^2=26.65 \GeV^2$, in  Fig.~\ref{fig:error_stat_config}.
In the histogram, the height of each bin, $N_i$ for the $i$-th bin, represents the number of counts for a shot with $\hat \p^2$ that falls into the corresponding interval.
Since each shot is independent, the probability distribution has a  Poisson form. We take its standard deviation, $\sigma_{St,i}=\sqrt{N_i}$, to be the statistical uncertainty for each bin, denoted by the red error bars in the figure. 
The configuration uncertainties, indicated by the blue error bars in the figure, are calculated as $\sigma_{\rho, i}= \sqrt{\sum_{I=1}^{K=5} (N_{I, i}-\overline N_i)^2/(K-1) }$ with 
$\overline N_i=\sum_{I=1}^K N_{I, i}/K$, where the index $I$ denotes the medium configurations.
We can see that with this number of shots, the statistical uncertainty is negligible, with a bin-averaged relative uncertainty of 1.2\%, especially when compared to the larger configuration uncertainties.
 In addition, for other setups, we observe that the statistical uncertainty is negligible with 819200 shots. In the simulation results that follow, we take this number of shots for each measurement and the plotted uncertainties are only from the medium configuration fluctuations.

\subsection{Momentum broadening}
\label{sec:res_U1}

With the jet state obtained from the quantum simulation, we are able to calculate the quenching parameter $\hat q$ defined in \eqn{eq:def_qhat}, as formulated in Sect.~\ref{sec:method_measure}. We perform the measurement by measuring all the qubits in the computer in each run of the simulation, thus reconstructing the underlying distribution.

To have a baseline to compare our numerical results, we first outline the analytical computation of the quenching parameter. The analytical derivation of $\hat q$ for a single quark jet can be reduced to the computation of Wilson line correlators, assuming the eikonal limit and that the medium is homogeneous. Here, we briefly revisit the calculation in the discretized basis introduced in Sect.~\ref{sec:basis_encoding}. For that, we first write \eqn{eq:def_qhat} explicitly as
\begin{align}\label{eq:qhat_version2}
\hat q  =\frac{1}{L_\eta} \llangle \bra{\psi_0} U^\dagger(L_\eta;0) \hat \p^2  U(L_\eta;0) \ket{\psi_0} \rrangle \, ,   
\end{align}
where the initial state $\ket{\psi_0}$ describes a state with $\p=0$, also used in the simulations. In coordinate space, \eqn{eq:qhat_version2} reduces to 
\begin{align}\label{eq:qhat_def_cont}
 \hat q =     \frac{1}{L_\eta} \int_{\x,\y,\k} e^{-i\k \cdot  (\y-\x)} \k^2 \llangle \cW^\dagger(\y) \cW(\x) \rrangle \, ,
\end{align}
where $\cW$ is a light-like Wilson line along $x^+$
\begin{align}
\cW(\x)= \exp \left(-ig \int_{0}^{L_\eta} dx^+ A^-(x^+,\x)\right) \, .   
\end{align}
Notice that $\cW$ is nothing but the time evolution operator given by \eqn{eq:master_eq} in the exact eikonal limit, and it describes the multiple gluons exchanges between the jet probe and the medium.\footnote{It has been shown that the inclusion of the kinetic operator at finite $p^+$ and at leading eikonal order does not affect $\hat q$,  cf. Ref.~\cite{Blaizot:2012fh}.} 
Then taking the correlation relation \eqn{eq:chgcor}, \eqn{eq:qhat_def_cont} can be computed on the discrete transverse lattice, and the result is
\begin{multline}\label{eq:qhat_analytic}
\hat q =    \frac{C_F g^4\mu^2}{4\pi}
\\
\times
\left[ \log \left(1+\frac{ \pi^2}{a_\perp^2m_g^2}\right)- \frac{1}{1+a_\perp^2 m_g^2/\pi^2} \right]  \, , 
\end{multline}
where the Coulomb logarithm emerges due to the continuum UV divergence, here regulated by the ratio between the largest momentum mode in the lattice, $\pi/a_\perp$, and the IR regulator $m_g$.

\begin{figure}[t]
    \centering
    \includegraphics[width=0.4\textwidth]{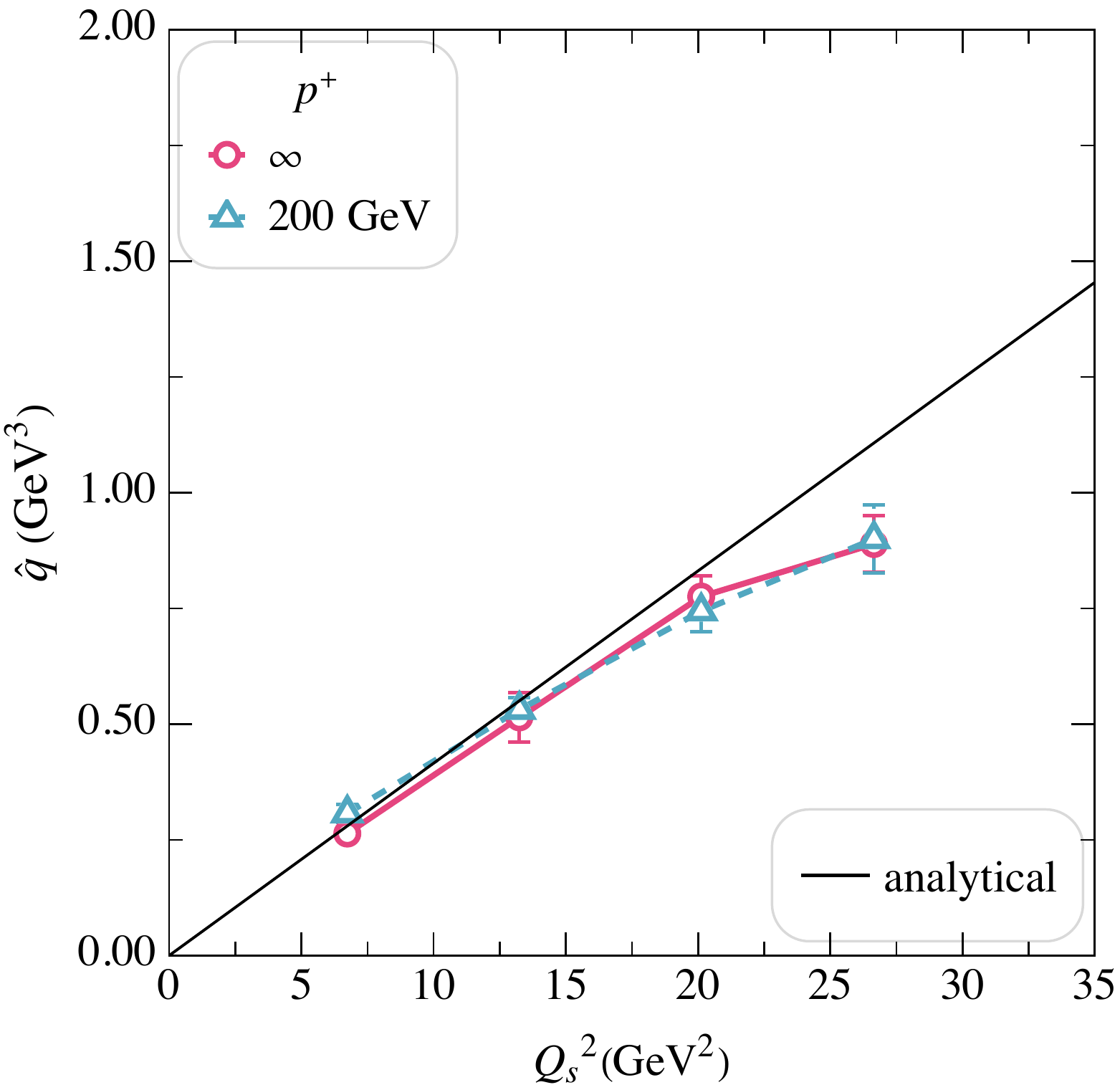}
    \caption{
    Quenching parameter $\hat{q}$ in a U(1) medium as a function of the saturation scale $Q_s^2$. Results at finite (blue triangles) and infinite (red circles) jet energy are compared to the analytical result given in Eq.~\eqref{eq:qhat_analytic}. Parameters used in the simulations: $L_\perp=4.8 \GeV^{-1}$,  $m_g=0.8 \GeV$, and $N_\perp=16$.}
    \label{fig:results_u1}
\end{figure}

Let us first consider the case of a U(1) homogeneous medium. 
In Fig.~\ref{fig:results_u1} we show the extracted values of the jet quenching parameter $\hat{q}$ as a function of the saturation scale $Q_s^2$ at selected values of $6.73$, $13.24$, $20.12$, and $26.65  \GeV^2$. We run the simulations on the basis of $N_\perp=16$, such that the lattice spacing is $a_\perp=0.3 \GeV^{-1}$. We consider both the eikonal ($p^+=\infty$) and subeikonal  ($p^+=200 \GeV$) cases: the former  are shown in red circles while the latter are denoted by blue open triangles. Each data point is averaged over five medium configurations, and the error bars are calculated as the standard deviation.
The analytical result given in \eqn{eq:qhat_analytic} is shown in the solid black line for comparison.

We would like to address two key observations from here.
First, the two sets of results at infinite and finite $p^+$ overlap within their uncertainties. This agreement indicates that the kinetic energy term $\hat K$ does not contribute to $\hat q$, as expected. Note that the uncertainty bar tends to increase with growing $Q_s^2$. This is due to the fact that the fluctuations of the medium are enhanced when the sources admit a larger variance ($\sim g^2\mu$) as a random Gaussian variable [see also \eqn{eq:f_rho}].

Second, the results from the simulations agree with the analytical form in \eqn{eq:qhat_analytic} in the lower $Q_s^2$ regime but start to deviate at larger $Q_s^2$. This deviation results from the underlying transverse lattice being finite. When the jet state reaches the boundaries of the lattice, the momentum square $\p^2$ can no longer increase linearly as one would expect from the analytical derivation of \eqn{eq:qhat_analytic}, but instead, results in the observed nonlinear behavior.

\begin{figure*}[]
    \centering
    \subfigure[\label{fig:qhat_aperp} ]{
    \raisebox{0.1\height}{ \includegraphics[width=0.4\textwidth]{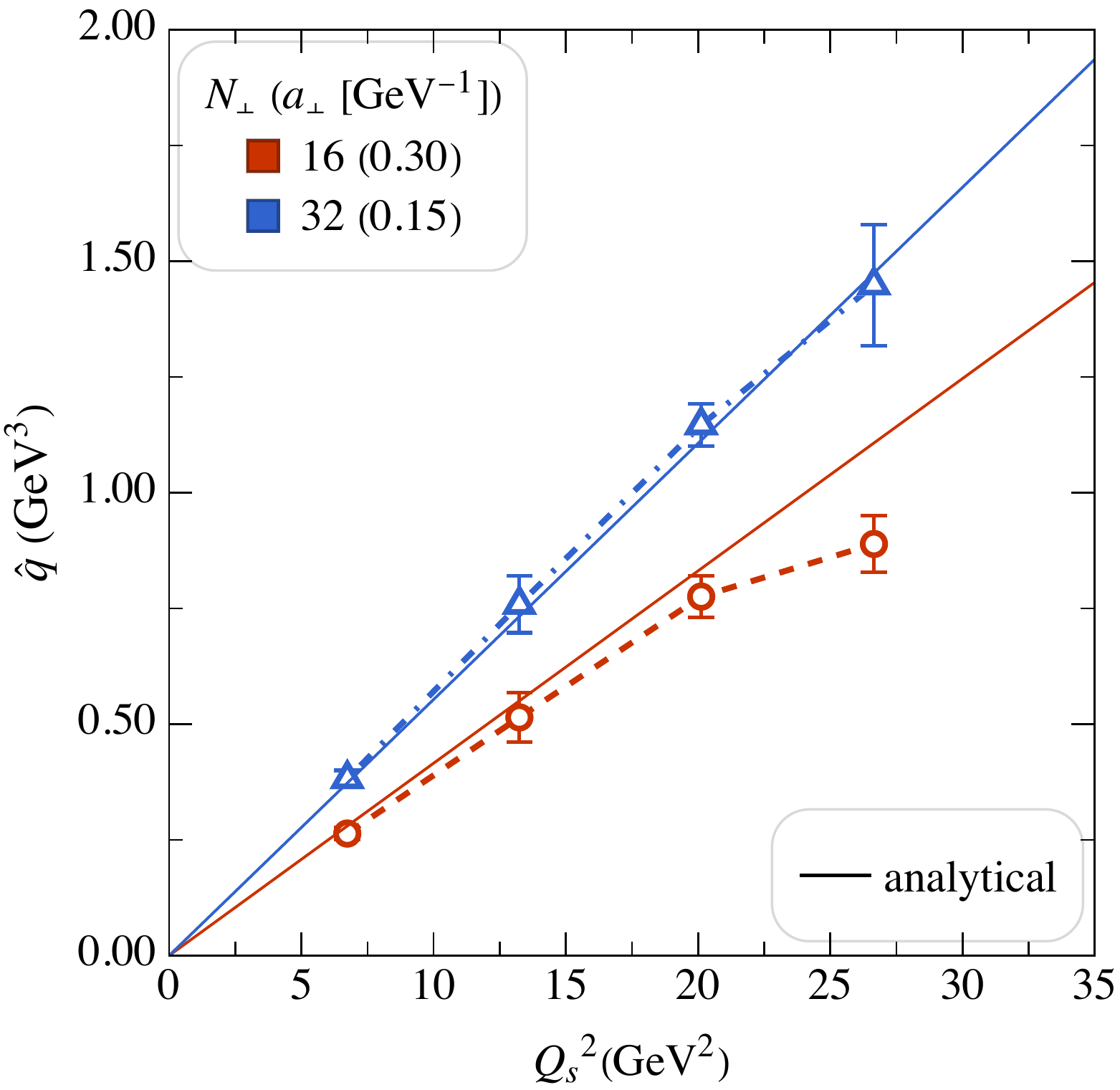}
    }}
    \quad
    \subfigure[ \label{fig:qhat_aperp_2D} ]{
    \includegraphics[width=0.5\textwidth]{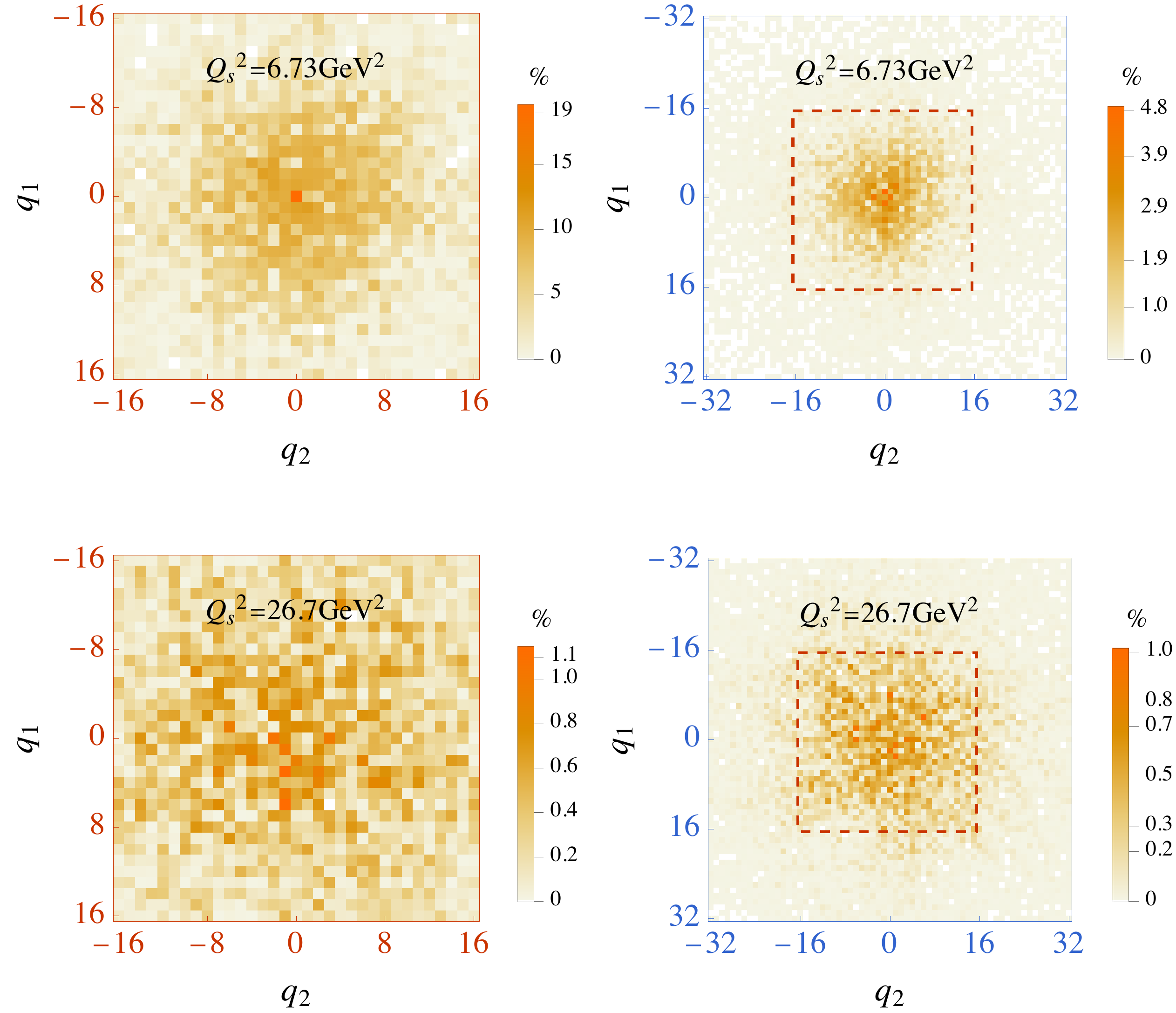}}
    \caption{
    A comparison of simulations between the lattices $N_\perp=16$ and 32 at $L_\perp=4.8 \GeV^{-1}$. (a) Quenching parameter $\hat{q}$ as a function of the saturation scale $Q_s^2$. (b) Transverse momentum distribution of final state at $Q_s^2=6.73 \GeV^2$ and $Q_s^2=26.65 \GeV^2$. In (b), the range of right panel's lattices is four times as large as the left ones. The dashed squares indicate the region covered by the smaller lattices. 
    }
    \label{fig:results_2d_plots_U1}
\end{figure*}

To further examine the deviation at large $Q_s^2$, we run the simulations with a $\p$-larger lattice, using $N_\perp=32$, such that the lattice spacing is halved $a_\perp=0.15 \GeV^{-1}$ and the covered momentum range is doubled $(\lambda_{UV}=\pi/a_\perp)$, taking $p^+=\infty$ case and without changing any other parameters. The results are shown in Fig.~\ref{fig:results_2d_plots_U1}. 
In Fig.~\ref{fig:qhat_aperp}, the quenching parameter $\hat{q}$ at various saturation scales are shown in red circles (blue open triangles) for the results at $N_\perp=16$ (32), and the analytical results according to \eqn{eq:qhat_analytic} are shown in solid lines with their respective colors. Note that both the simulation results and the analytics have a steeper slope at $N_\perp=32$ compared to $N_\perp=16$, resulting from having a finer resolution in probing the UV-divergent medium [see also \eqn{eq:chgcor}]. 
From the figure, one can see that the simulation results at $N_\perp=32$ agree well with the analytical expectation even at larger $Q_s^2$, in comparison to $N_\perp=16$.

In addition, we present in Fig.~\ref{fig:qhat_aperp_2D} the transverse momentum distribution of the final jet state at the smallest and largest values of the saturation scale $Q_s^2$ shown in Fig.~\ref{fig:qhat_aperp}, for both lattices considered. The results are shown for a single medium configuration.
Indeed, at the smaller saturation scale of $Q_s^2$=6.73\,GeV$^2$ (the top panels), one can see that in both lattices, a large part of the state is distributed away from the boundaries. On the contrary, with a stronger medium with the larger saturation scale $Q_s^2$=26.65\,GeV$^2$ (the bottom panels), the state largely reaches the boundaries on the $\p$-smaller lattice ($N_{\perp}$=16, left panel), whereas it is still further away from the boundaries on the $\p$-larger lattice ($N_{\perp}$=32, right panel). In other words, at a large saturation scale, the smaller lattice becomes ``saturated" by the state.

The above simulation results have helped verify our quantum simulation algorithm and examine the lattice effects entering the observable. It should be noted that although increasing the lattice size from $N_{\perp}$=16 to 32 could potentially mitigate the lattice effects, it requires one extra qubit [see \eqn{eq:nQ}], making it computationally more demanding when performing simulations in more complicated cases, such as at finite $p^+$. For this reason, in the following results we mostly use the lattice with $N_{\perp}$=16, being aware that the lattice effect enters the observable, especially at large $Q_s^2$.

\subsection{Momentum broadening with color rotations }
\label{sec:res_SU2}

\begin{figure}[t]
    \centering
\includegraphics[width=0.4\textwidth]{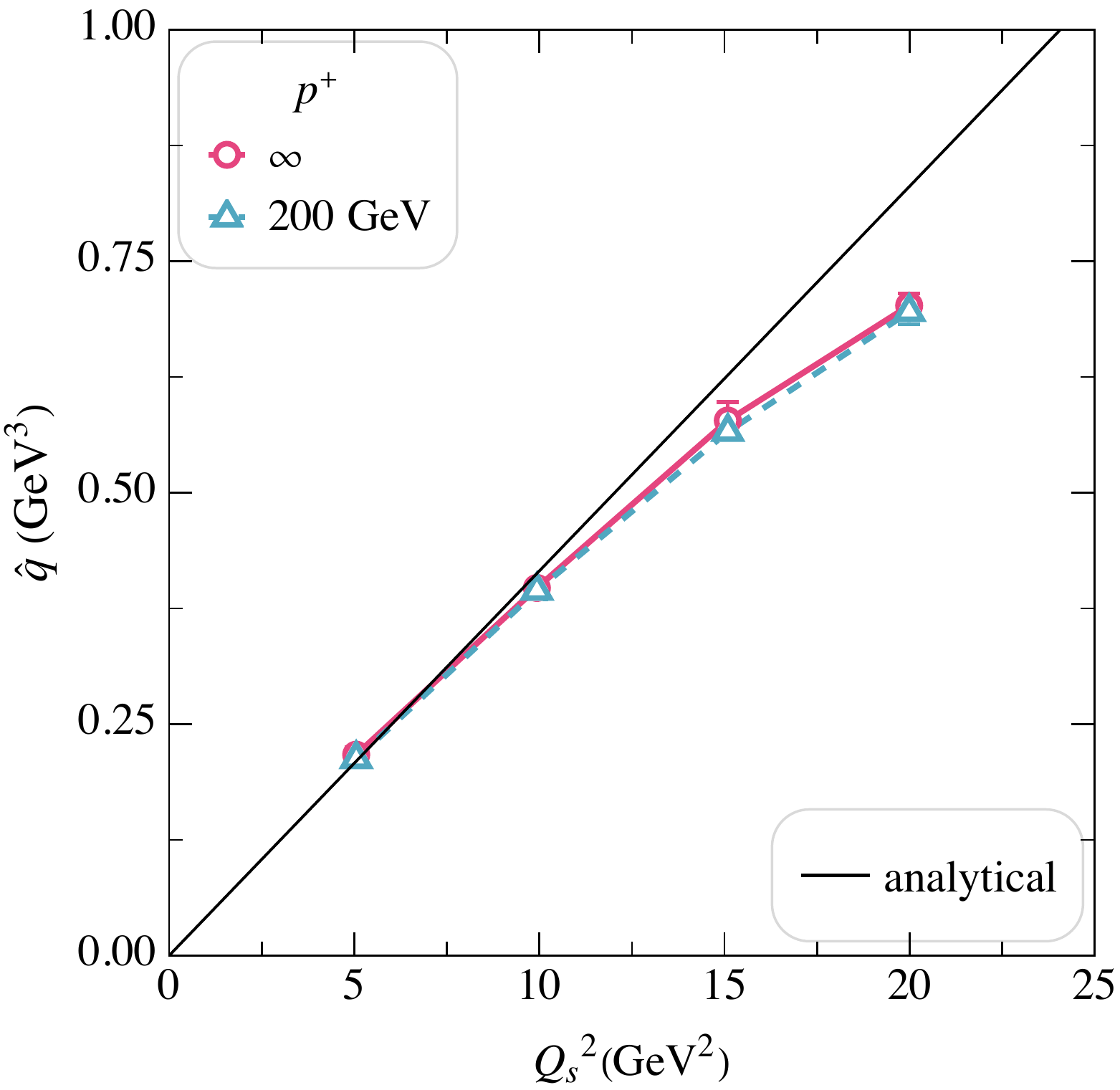}
    \caption{
    Quenching parameter $\hat{q}$ in a SU(2) medium as a function of the saturation scale $Q_s^2$. The parameters used in the simulations are the same as in Fig.~\ref{fig:results_u1}. The analytical result is given by Eq.~\eqref{eq:qhat_analytic}.
    }
    \label{fig:results_su2}
\end{figure}

Having studied the jet momentum broadening with a U(1) medium, we now introduce a SU(2) color dimension. 
 We present the results of the quenching parameter $\hat q$ at various saturation scales $Q_s^2$ in Fig.~\ref{fig:results_su2}. 
The parameters in these simulations are the same as those in Fig.~\ref{fig:results_u1}, with the color degree of freedom requiring one additional qubit. Note that with the color dimension, the values of $Q^2_s$ slightly vary since we keep the array of $g^2\mu$ the same as before, such that $Q^2_s=$5.05, 9.93, 15.09, and $20.00 \GeV^2$.
As we have observed in the U(1) case, in Fig.~\ref{fig:results_u1}, the two sets of results at infinite and finite $p^+$ agree. It is therefore further verified that the quenching parameter $\hat q$ in terms of $Q_s^2$ is not sensitive to the kinetic energy term.
The comparison to the analytical result given by \eqn{eq:qhat_analytic} also suggests an overall agreement with a deviation at increasing $Q_s^2$. The deviation happens due to the lattice effect already discussed in Sect.~\ref{sec:res_U1}.
We also measure the jet's color differential and total transverse momentum distributions, for which we place selected results in App.~\ref{app:pperp_color} for interested readers.

\begin{figure}[t]
    \centering
    \includegraphics[width=0.4\textwidth]{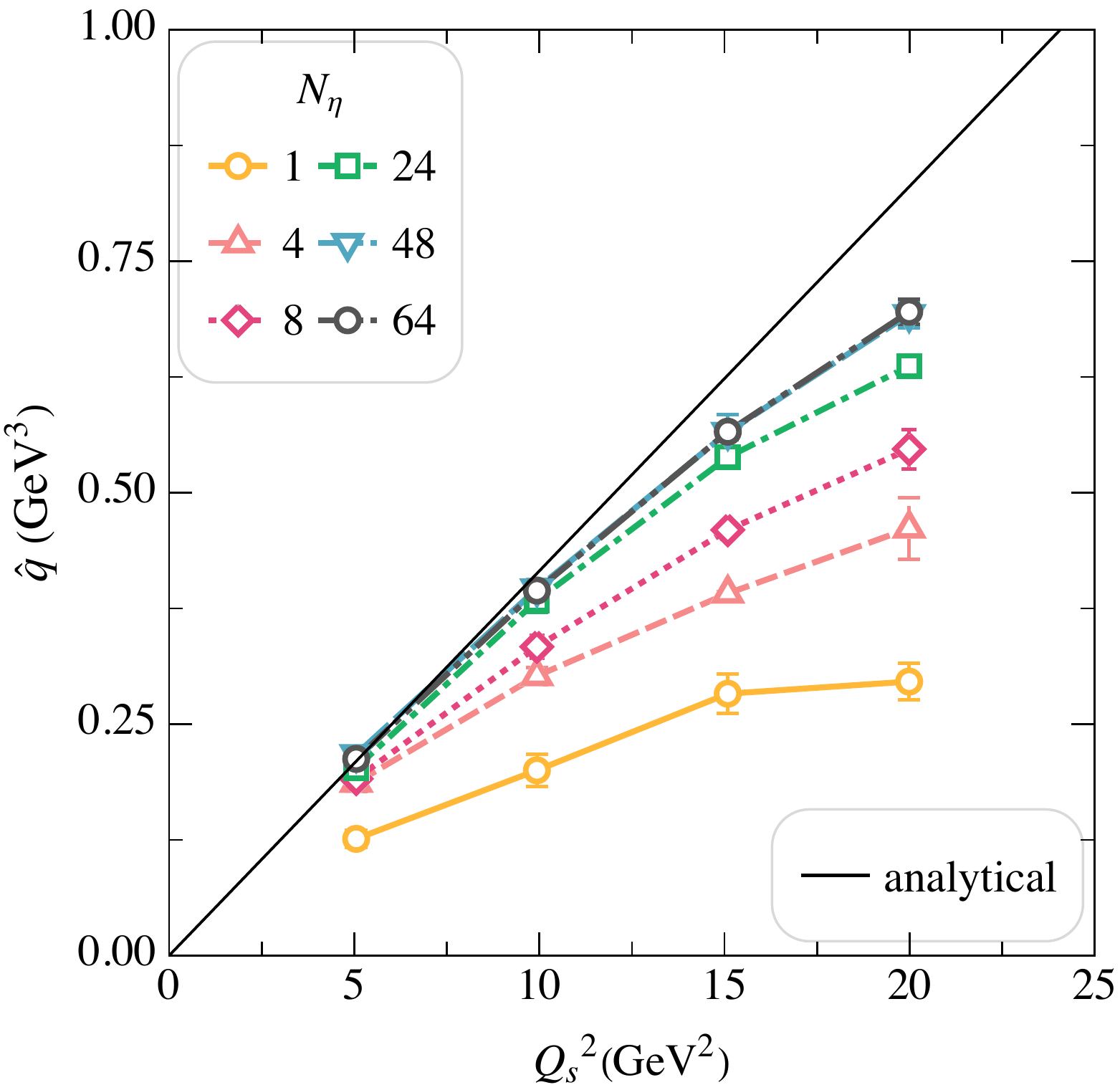}
    \caption{
    Quenching parameter $\hat{q}$ for a SU(2) medium as a function of the saturation scale $Q_s^2$ for various number of layers $N_\eta$. The jet energy is fixed at $p^+=200 \GeV$. The data points at $N_\eta=64$ are the same as those in Fig.~\ref{fig:results_su2}. The remaining simulation parameters are the same as used in Fig.~\ref{fig:results_u1}. The analytical result is given by Eq.~\eqref{eq:qhat_analytic}.
    }
    \label{fig:layers}
\end{figure}

In the simulations, we treat the medium as having multiple layers along $x^+$, as formulated in Sect.~\ref{sec:gate encoding}. In particular, we take $N_\eta=64$ in producing the presented results. 
We would like to emphasize the necessity of using multiple layers, by demonstrating the evolution of a probe with $p^+=200 \GeV$ in the SU(2) background, with various $N_\eta$.
In Fig.~\ref{fig:layers}, we show $\hat q$ at various $Q_s^2$ extracted from simulations using  increasing values for $N_\eta$. With just a single layer($N_\eta=1$), there is a sizable discrepancy between the simulation and the analytical results. However, as the number of layers increases, we find that all points at different $Q_s^2$ converge towards the large $N_\eta$ result, which agrees with the analytical result up to lattice effects. In addition, the result at $N_\eta=48$ already overlaps with that at $N_\eta=64$, indicating that the latter parameter is sufficient to capture the longitudinal structure of the medium for our simulations. 
The need of introducing layers into the quantum simulation makes it distinct from similar approaches in, for example, quantum chemistry, where the potential term is not stochastic; for the latter, see e.g., Ref.~\cite{Kassal_2008}.

\subsection{Momentum broadening in anisotropic media
}
\label{sec:res_anisotropic}


In recent times, there has been an increase interest in studying jet evolution in the presence of anisotropic backgrounds~\cite{Barata:2022krd,Fu:2022idl,Sadofyev:2021ohn}. Such studies aim to provide a better understanding of how jets can be used as tomographic tools of the medium~\cite{He:2020iow,Vitev:2002pf}
Here we extend the previous results and exemplify how our approach can be used to study jet evolution in structured matter. 

In particular, following \eqn{eq:poisson} and \eqn{eq:chgcor}, we consider medium profiles 
where either the IR regulator $m_g$ or the medium density $g^2\mu$ have a spatial dependence. For the former case we use the fact that \eqn{eq:poisson} is formally solved for a constant $m_g$ by \eqn{eq:MVA_Green_p1} and then reinstate a transverse coordinate dependence on the thermal mass, such that 
\begin{align}\label{eq:MVA_Green_p1}
  A_a^-( x^+,\x)=\int_{\z,\k}  \frac{e^{-i \k \cdot (\x-\z)}}{ m_g^2(\x)+\k^2} \rho_a( x^+,\z) \; .
\end{align}
Following the strategy used in Refs.~\cite{Barata:2022krd,Sadofyev:2021ohn}, we allow $m_g$ to vary linearly along a direction in transverse space,
\begin{align}\label{eq:mg_r}
    m_{g}\to m_g^2(\x)=m_{g}^2(1+\c\cdot \Delta \x)
  \, ,
\end{align}
where in spatial coordinates we have that $a_ \perp \c=(\nabla m_g,0)^T$ and $\Delta \x = \x- (-L_\perp,0)^T$. 
The anisotropy effects are controlled by the parameter $\nabla m_g$. In the following numerical results we took $\nabla m_g= 0.01, 0.10$. 

Anisotropic effects can also be included by allowing the background field to have an anisotropic profile in transverse space. For that we follow the strategy used in Ref.~\cite{Li:2020uhl} and apply a profile function on the field $A^-_a$,
\begin{align}
  A_a^-( x^+,\x)\to A_a^-( x^+,\x) f(\x)\; .
\end{align}
We consider the profile function in the format of
\begin{equation}\label{eq:f_r}
f(\x) = 1+ \c\cdot \Delta \x 
   \;,
\end{equation}
where $a_\perp \c =(\nabla f,0)^T$ and $\nabla f=0.01,0.1 $. 
This smearing of the background field tries to capture the spacetime dependence of quasi particles in the plasma, and its effect is phenomenologically similar to modifying the charge density $g^2\mu$.

\begin{figure}
    \centering
    \subfigure[\ $\nabla m_g$ effect]{\includegraphics[width=0.4\textwidth]{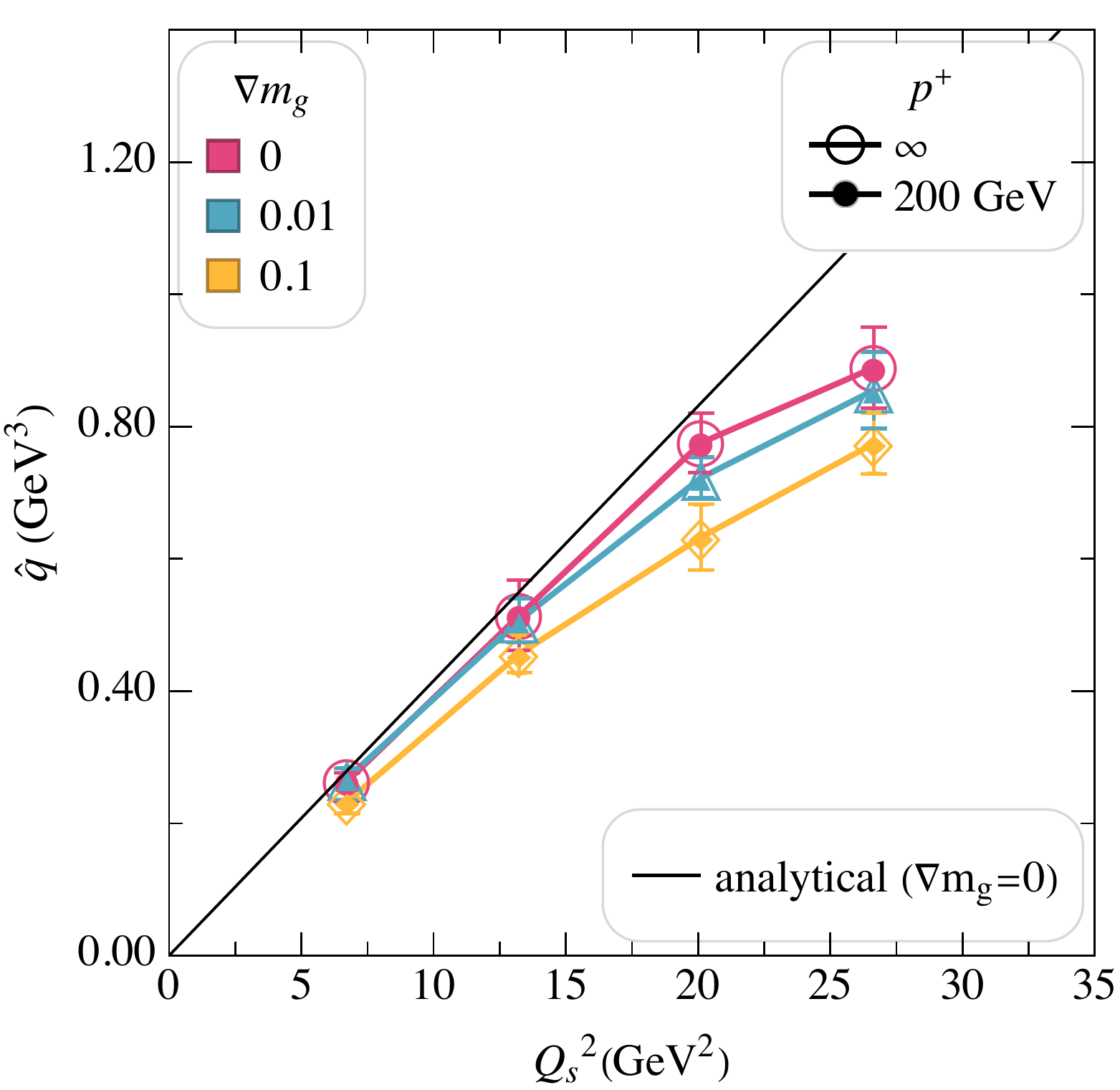}}
    \subfigure[\ $\nabla f$ effect]{\includegraphics[width=0.4\textwidth]{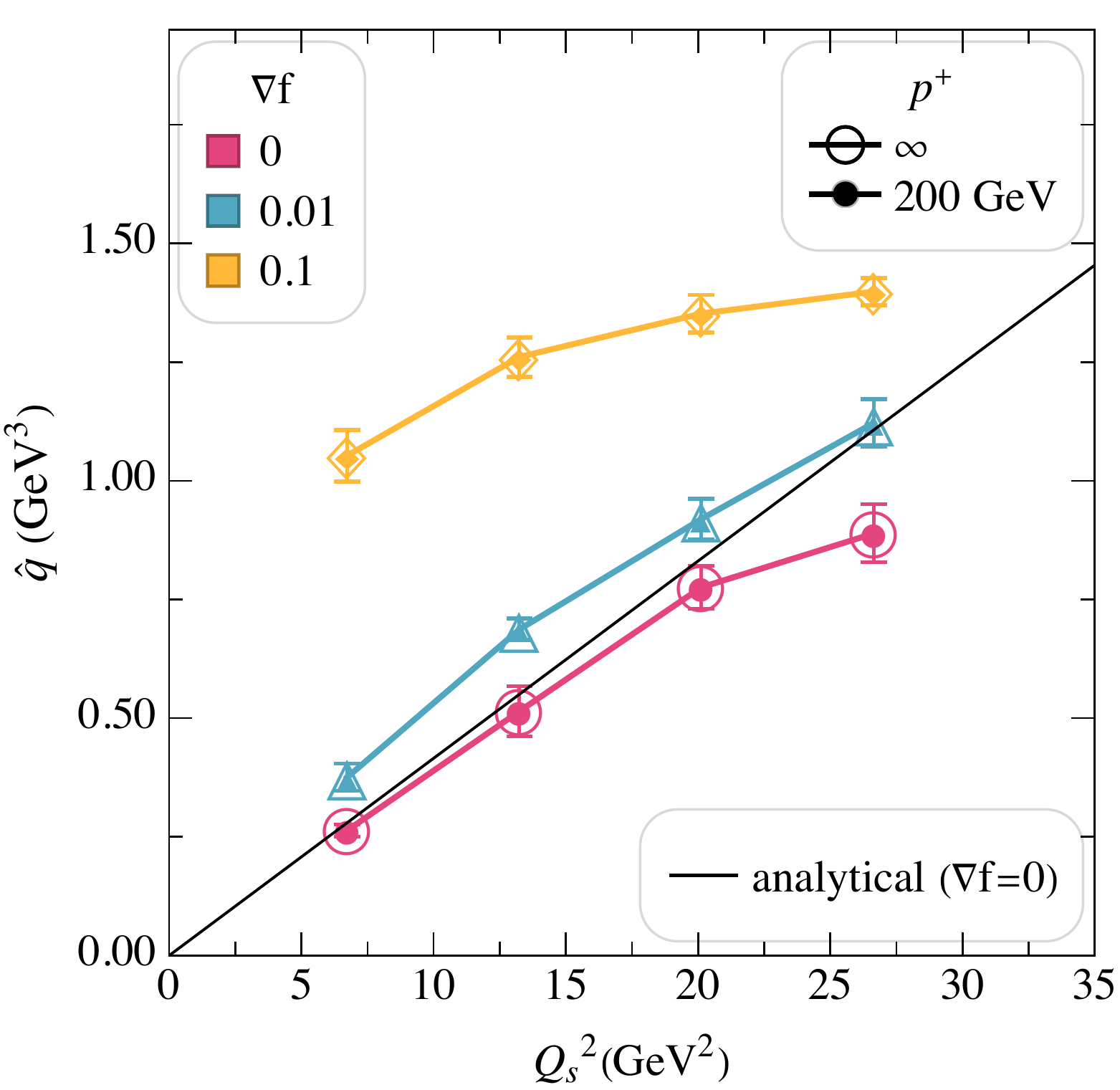}}
    \caption{
   Quenching parameter $\hat{q}$ in a U(1) medium as a function of the saturation scale $Q_s^2$, including anisotropic effects in the screen mass (a) and the medium density (b). Results at finite (infinite) jet energy are shown with closed (open) markers. The parameters used are the same as in Fig.~\ref{fig:results_u1} (a). The analytical result is given by Eq.~\eqref{eq:qhat_analytic}.
    }
    \label{fig:mgCx}
\end{figure}

In Fig.~\ref{fig:mgCx}, we present the simulation results using $N_\perp=16$ and considering a U(1) background, including $\nabla m_g$ (top) and  $\nabla f$ (bottom) effects. The curves without anisotropic effects ($\nabla f=\nabla m_g=0$) match the ones shown in Fig.~\ref{fig:results_u1}. 
For both types of modifications, we observe that there is a clear sensitivity of $\hat q$ to the medium profile. This is especially true for $\nabla f $ effect. Such corrections are expected to get possible logarithmic enhancements compared to the modification of the thermal mass~\cite{Barata:2022krd}. Interestingly, we see that the anisotropy effect alters $\hat q$ differently in both scenarios: for $\nabla m_g$ the jet suffers less broadening, while for $\nabla f$ we see a drastic increase in the diffusion coefficient. This is expected, because the former correction leads to an overall weaker field, whereas the latter a stronger field.

Another interesting observation comes from comparing simulations using the same background profile but at different jet energies $p^+$. Indeed, for all studied cases we see that the corrections to $\hat q$ are not sensitive to the kinetic phases. This result might appear to contradict the findings in  Ref.~\cite{Barata:2022krd}, where it was shown that (1) $\hat q$ gets no correction at leading order in the anisotropy coefficient and (2) all other possible corrections enter only at sub-eikonal accuracy, and thus should be highly sensitive to $p^+$. In this work, $\hat q$ receives contributions from all power corrections in the anisotropy coefficient through direct exponentiation in the time evolution. Therefore, it captures both the energy dependent corrections considered in Ref.~\cite{Barata:2022krd}, and higher order terms, which can become dominant. A possible way to explore the same region as in Ref.~\cite{Barata:2022krd} is to consider a much smaller jet energy $p^+$ and calculate directional expectation values such as $\langle \hat\p \,\hat\p^2\rangle $.

\subsection{
Simulations in a noisy quantum computer}
\label{sec:res_noise}

When running the circuit in a real device, the simulations become sensitive to quantum errors and decoherence effects in the quantum computer.
In addition, the current circuit implementation has not been optimized to reduce the impact of such effects on the extracted results.
We leave such improvements for a future study.

\begin{figure}[t]
    \centering
    \subfigure{\includegraphics[width=0.4\textwidth]{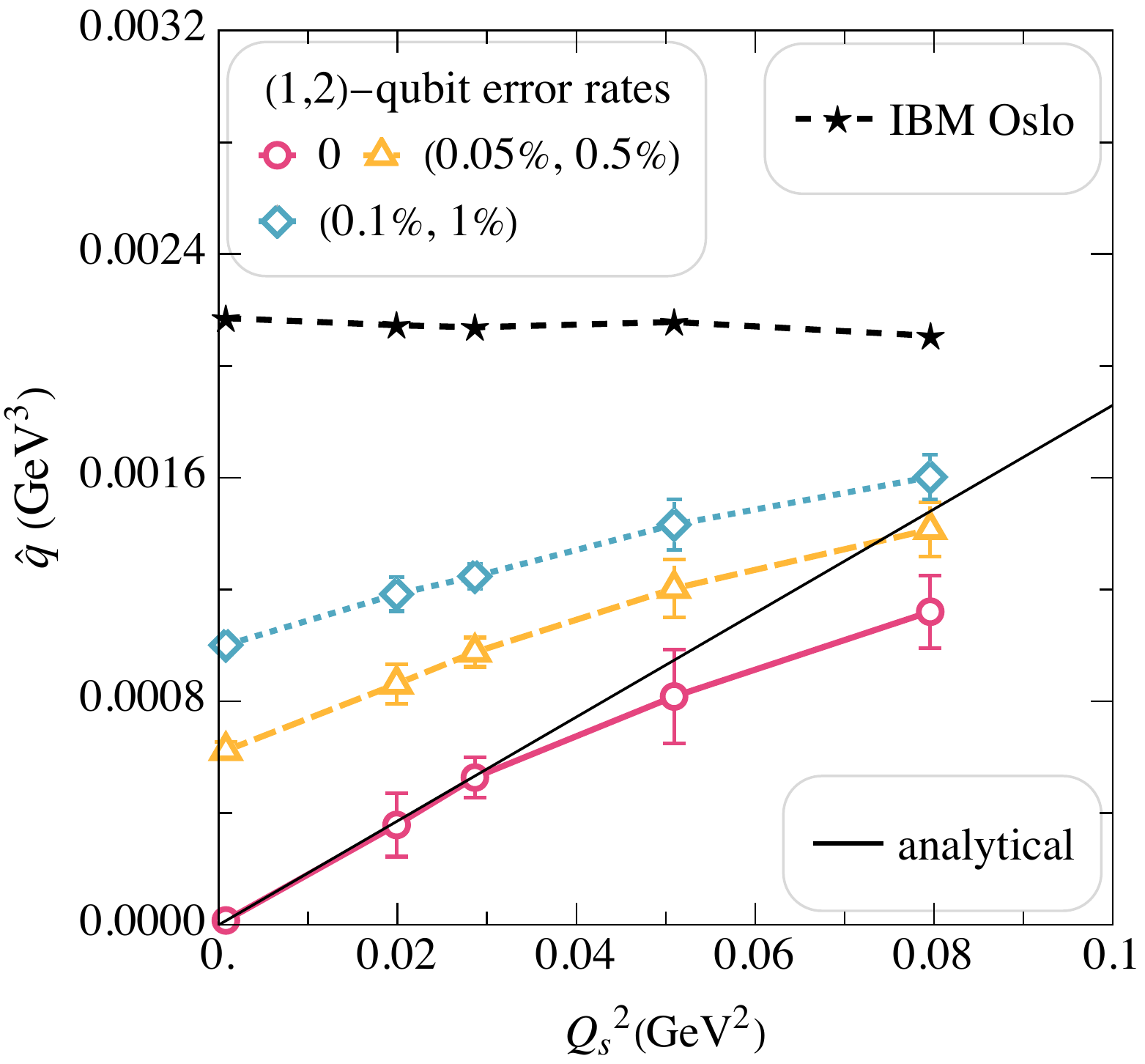}}
    \caption{   
      Quenching parameter $\hat{q}$ in a U(1) medium as a function of the saturation scale $Q_s^2$, extracted from noisy simulations (open markers) and the {\tt IBM Oslo} quantum computer (black stars) at infinite jet energy. For the {\tt IBM Oslo} data,  each data point is measured with one medium configuration using the maximal available number of shots: 20000. Other parameters used in the simulations: $N_\perp=4$, $L_\perp=32 \GeV^{-1}$,  $m_g=0.1 \GeV$, $L_\eta=50 \GeV^{-1}$, and $N_\eta=1$. 
      The analytical result is given by Eq.~\eqref{eq:qhat_analytic}.
    }
    \label{fig:noise}
\end{figure}

In Fig.~\ref{fig:noise}, we present our results from simulations performed using simple noise models, implemented using the QASM {\tt qiskit} backend, and the results from running the simulation in a public IBM quantum processor.
The extracted quenching parameter $\hat q$ is plotted as a function of the saturation scale $Q_s^2$, in the case of a U(1) background, taking $N_\perp=4$, $a_\perp=8 \, {\rm GeV}^{-1}$ and $m_g=0.1$ GeV. Though this set of parameters is not expected to depict a real physical scenario, it requires a small computational time and its features replicate those observed for the previous results. 
The solid black curve provides the analytical baseline according to Eq.~\eqref{eq:qhat_analytic}, while the zero noise points give the result in the case of an ideal quantum computer. The red (green) data set provides the results for the case of quantum computer with 1-qubit error rates of $0.05\%$ ($0.1\%$) and 2-qubit error rates of $0.5\%$ ($1\%$). The error probabilities correspond to the value taken by the depolarizing error parameter in the {\tt qiskit} native method {\tt depolarizing\char`_error}.

Comparing all the curves we observe that indeed the current implementation is not resilient to even the simplest noise model. This reflects the fact that the circuit for the $U_A$ operator has not been optimized to reduce its length and gate complexity. This could be done by, for example, implementing the operators approximately (i.e., within some error bound of the exact quantum gate) or by discretizing the background field values~\cite{Kassal_2008}. We note however, that for both error models, the resulting curves are essentially shifted from the ideal one by a constant, leading us to speculate that the zero noise result might be extractable by extrapolation~\cite{Li_2017,Temme_2017}. We expect this trend to be modified when including more realistic noise models. Also note that the sensitivity to the computer noise will most likely worsen as $N_\perp$ increases.

For the simulation using a real device, the last set of data points depicts the output obtained from running the circuit 
in the recently-released public quantum computer {\tt IBM Oslo}. In addition to various single and double qubit errors (0.032~\% and 0.864~\% respectively), {\tt IBM Oslo} also experiences additional readout error averaged around 1.68~\% and a limited quantum volume of 32, which is expected to be insufficient for the current circuit implementation.
Indeed, one observes that the result is essentially dominated by noise and one can not extract physically relevant information from the output.

\section{Conclusion and outlook }
\label{sec:conclusion}

In this work, we have developed a framework to simulate medium induced jet broadening on a quantum computer. Our formulation is based on the light-front Hamiltonian formalism, particularly in the recent development of the tBLFQ approach~\cite{Li:2020uhl}, implementing the quantum algorithm proposed in Ref.~\cite{Barata:2021yri}. We numerically simulated the time evolution of a single quark jet, thus providing the opportunity to study effects beyond the eikonal limit and the evolution in more realistic media. 

In the current approach, the jet quantum state after having traversed the medium was prepared by implementing a digital quantum algorithm. By performing multiple projective measurements of such a state, we extracted the underlying transverse momentum distribution and the jet quenching parameter $\hat q$ at various saturation scales.
In this study we considered the jet evolution in homogeneous medium, for both the U(1) and SU(2) probes. We showed that the simulation results agree with the analytical ones. We then studied two types of anisotropic medium profiles, where the problem is hard to tackle out analytically. In both scenarios, we found sizable corrections to $\hat q$, revealing the importance of studying such medium configurations for jet quenching phenomenology.
Lastly, we examined the behavior of the constructed circuit to noisy quantum simulators, and to a real quantum processor. We found that the current circuits need to be further optimized to deal with quantum noise, which we leave for a future study.

The current approach can be extended to the case where gluon emissions are included. In such a regime, it is expected that quantum computers surpass their classical counterparts, since the problem's computational complexity scales exponentially with the number of particles.
In forthcoming work we will address such questions, extending the present algorithm to include soft radiation produced from the hard part of the jet. Such an application might provide further insight into radiative corrections to momentum broadening~\cite{Ghiglieri:2022gyv,Caucal:2021lgf,Caucal:2022fhc}, color coherence effects~\cite{Dominguez:2019ges,Casalderrey-Solana:2012evi} or the QCD LPM effect with multiple gluons~\cite{Arnold:2015qya}. On the other hand, the current approach can also be extended to jet evolution in other phases in heavy ion collisions, such as the glasma phase at early stages~\cite{Ipp:2020mjc,Carrington:2021dvw,Hauksson:2021okc}.

On the strict quantum circuit implementation there are several open challenges. The principal future task is to reduce the circuit depth, so that the computation can be more efficient and noise resilient.
We plan to improve the quantum algorithms for the time evolution operator and use approximate (instead of exact) implementations for the quantum gates, as mentioned in Sect.~\ref{sec:method_measure}. Another possible direction is to write the Hamiltonian in terms of Pauli strings, which might be convenient when looking at a specialized background fields in related problems. Finally, for the circuit implementation to ever work in a real device, error mitigation and error correction strategies have to be implemented, as discussed above.

\section*{Acknowledgments}
We are very grateful to Robert Basili, Miguel A. Escobedo, Tuomas Lappi, Andrey Sadofyev, James P. Vary, Xin-Nian Wang, Bin Wu, and Xingbo Zhao for helpful and valuable discussions. We acknowledge the use of IBM Quantum services for this work. The views expressed are those of the authors, and do not reflect the official policy or position of IBM or the IBM Quantum team. 
JB is supported by the U.S. Department of Energy, Office of Science, National Quantum Information Science Research Centers under the “Co-design Center for Quantum Advantage” award and by the U.S. Department of Energy, Office of Science, Office of Nuclear Physics, under contract No. DE-SC0012704. 
XD is supported by the Deutsche Forschungsgemeinschaft (DFG) under grant CRC-TR 211 “Strong-interaction matter under extreme conditions” project no.~315477589-TRR 211. 
ML, WQ, and CS are supported by Xunta de Galicia (Centro singular de investigacion de Galicia accreditation 2019-2022), European Union ERDF, the “Maria de Maeztu” Units of Excellence program under project CEX2020-001035-M, the Spanish Research State Agency under project PID2020-119632GB-I00, and European Research Council under project ERC-2018-ADG-835105 YoctoLHC.
WQ is in part supported by the U.S. Department of Energy (DOE) under Grant No. DE-FG02-87ER40371.

\appendix
\section{Conventions}
\label{appendix:conventions}
The light-front coordinates are defined as \( (x^+, \x, x^-) \), where \(x^+=x^0+ x^3\) is the light-front time,  \(x^-=x^0-x^3\) the longitudinal coordinate, and  \(\x=(x^1, x^2)\) the transverse coordinates. The letters in bold, such as $\x$, denote  transverse vectors, while their magnitude is denoted by $x_\perp\equiv |\x|$. 
The non-vanishing elements of the metric tensors $g^{\mu\nu}$ and $g_{\mu\nu}$ are, $ g^{+-}=g^{-+}=2$, $g_{+-}=g_{-+}=1/2$, $g^{ii}=g_{ii}=-1$ with $i=1,2$. 

In addition, we use the following shorthand notation
\begin{align}
\int_\x \equiv \int d^2\x \, , \quad     \int_\p \equiv \int \frac{d^2\p}{(2\pi)^2} \, , 
\end{align}
for transverse integrals in position and momentum space, respectively. Upon discretization, integrations over the phase space convert to sums over all lattice points, such that 
\begin{align}
\int_\x \to a_\perp^2 \sum_\x \, , \quad   \int_\p \to \quad b_\perp^2\sum_\p  \, .
\end{align}

In the continuum, any state in the transverse plane can be written in terms of position of momentum space states $\ket{\x}$ and $\ket{\k}$, respectively. These two basis are related by a Fourier transform, which when discretized reads
\begin{align}
\ket{\k}\equiv \int_\x \, e^{-i \p\cdot \x} \ket{\x}= a_\perp^2 \sum_\n e^{- i \pi \n\cdot \q/N_\perp} \ket{\n \, a_\perp}  \, ,  
\end{align}
in agreement with the conventions used in Fourier transform definition in~\eqn{eq:FT}.

\section{Computation of the background field}~\label{app:field}
In this appendix, we briefly detail the classical computation of the background field, following the approach in, e.g., Ref.~\cite{Li:2020uhl}.

Formally, the solution to \eqn{eq:poisson} can be written as
\begin{align}\label{eq:MVA_Green_p1}
  A_a^-( x^+,\x)=\int_{\z,\k}  \frac{e^{-i \k \cdot (\x-\z)}}{m_g^2+\k^2} \rho_a( x^+,\z) \; .
\end{align}
Thus, given a color source $\rho_a$, we numerically solve \eqn{eq:MVA_Green_p1} over the transverse lattice. To insure that \eqn{eq:chgcor} is satisfied, we sample the color sources from the Gaussian distribution functional
\begin{align}\label{eq:f_rho}
    f [\rho_a(\x,x^+)]=N \exp \bigg[-\frac{\Delta x^+ (\Delta \x)^2}{g^2 \mu^2}  \rho^2_a(\x,x^+)\bigg]\;,
\end{align}
with $N$ a normalization constant, $\Delta x^+=\tau$ and $\Delta \x=a_\perp$, the smallest lengths that can be resolved by the delta functions in \eqn{eq:chgcor}. 

\section{Criteria for choosing simulation parameters}\label{app:criteria}
The discretized transverse basis used in the main text, introduces natural IR and UV cutoffs $\lambda_{\rm IR}=\pi/L_\perp$ and $\lambda_{\rm UV}=\pi/a_\perp=N_\perp\lambda_{IR}$.
To ensure the results are not sensitive to the discretization employed, we use two criteria: one ensuring the physics of interest is being captured (range coverage) and another ensuring that the result is not sensitive to the finiteness of the lattice (broadening coverage). 
    \begin{itemize}
        \item[(a)] \textbf{Range coverage} \\
    To ensure the physical range of interest is covered by the lattice, one should require $\lambda_{\rm IR}$ to be smaller than the physical IR regulator $m_g$, while $\lambda_{\rm UV}$ must be taken larger than typical momentum transfer $Q_s$. In terms of $a_\perp$, these conditions imply 
     \begin{align}\label{eq:range_coverage}
   \frac{\pi}{N_\perp m_g} \ll a_\perp \ll \frac{\pi}{Q_s}
         \;.
     \end{align}
\item[(b)] \textbf{Broadening coverage}\\
The evolution of the quantum state should not be sensitive to lattice edge effects otherwise the final distribution would become asymptotically uniform  due to the lattice periodicity. For the particular case of $\hat \p^2$, one would obtain 
\begin{align}\label{eq:pperp_asy}
\begin{split}
\braket{\p^2}\stackrel{L_\eta\gg 0}{\longrightarrow} &
  \frac{1}{(2N_\perp)^2}\sum_{i=-N_\perp}^{N_\perp-1}\sum_{j=-N_\perp}^{N_\perp-1}(i^2+j^2)
  b_\perp^2\\
  \approx &\frac{2}{3}\frac{\pi^2}{a_\perp^2}\equiv \braket{\p^2}_{\mathrm{asy}}
  \; .
  \end{split}
\end{align} 
This saturated expectation value is reached in a time $t_{sat.}\equiv \braket{\p^2}_{\mathrm{asy}}/\hat q $.
To avoid this type of edge effects, we require $ t_{sat.} > L_\eta$. In terms of $a_\perp$, this is equivalent to ensuring that
\begin{align}\label{eq:broadening_coverage_Lambda}
    \begin{split}
    \Lambda& ( a_\perp)\equiv- a_\perp + \frac{2\pi}{\sqrt{3}Q_s}\times
    \\
    & \quad\biggl[
    \log\left(\frac{1}{a_\perp^2 m_g^2/\pi^2}+1\right)
    -\frac{1}{1+a_\perp^2  m_g^2/\pi^2}
    \biggr]^{-1/2}
    \; ,
    \end{split}
\end{align}
\end{itemize}   
is always positive. In obtaining the above equation we used Eq.~\eqref{eq:qhat_analytic}.
     
\begin{figure}[H]
    \centering
    \includegraphics[width=0.4\textwidth]{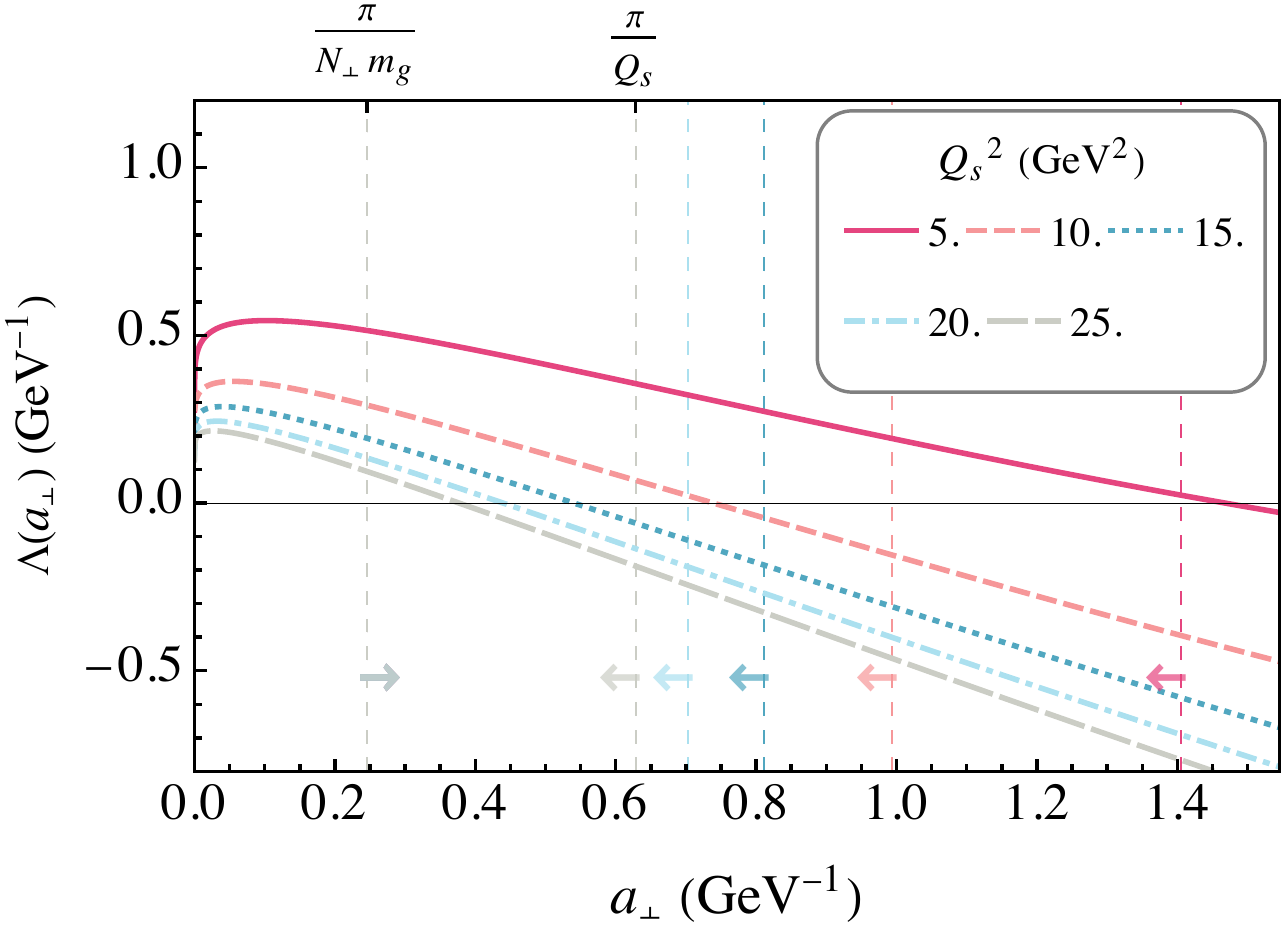}
    \caption{
    Criteria for selecting $a_\perp$ for various values of $Q_s$ and using $N_\perp=16$. The vertical dashed lines with arrows indicate the domain satisfying \eqn{eq:range_coverage} for each value of $Q_s$.
    Positive values of the $\Lambda$ curves indicate that the broadening coverage condition 
    is satisfied.
    }
    \label{fig:aperp_criteria}
\end{figure}

We illustrate the selection of $a_\perp$ using the two conditions discussed above, for the simulations shown in this work. 
Given the lattice of $N_\perp=16$, we consider the saturation scale $Q_s^2$ in the range of $5- 25 \GeV^2$, and set $m_g=0.8 \GeV$. We show in Fig.~\ref{fig:aperp_criteria}, the range coverage condition by dashed vertical lines denoting the domains satisfying \eqn{eq:range_coverage}, and the broadening coverage condition by plotting $\Lambda$ as a function of $a_\perp$.
We observe that the value $a_\perp=0.3 \GeV^{-1}$ is an eligible choice for this set up.

\section{Color differential measurement}\label{app:pperp_color}
\begin{figure*}[]
    \centering
    \subfigure[  ]{
    \includegraphics[width=0.78\textwidth]{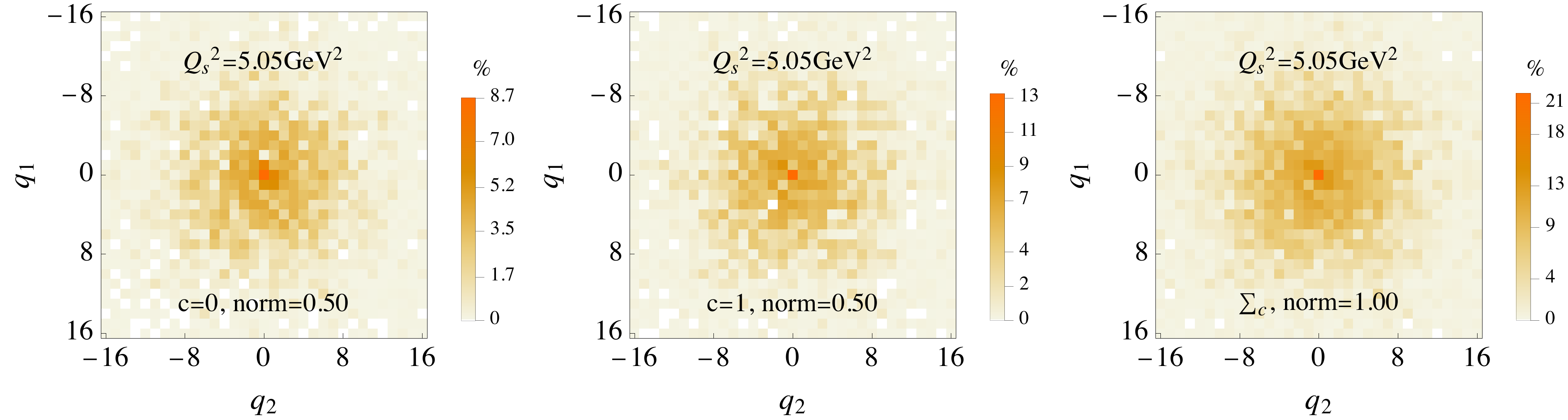}
  }
      \subfigure[  ]{
          \includegraphics[width=0.78\textwidth]{ 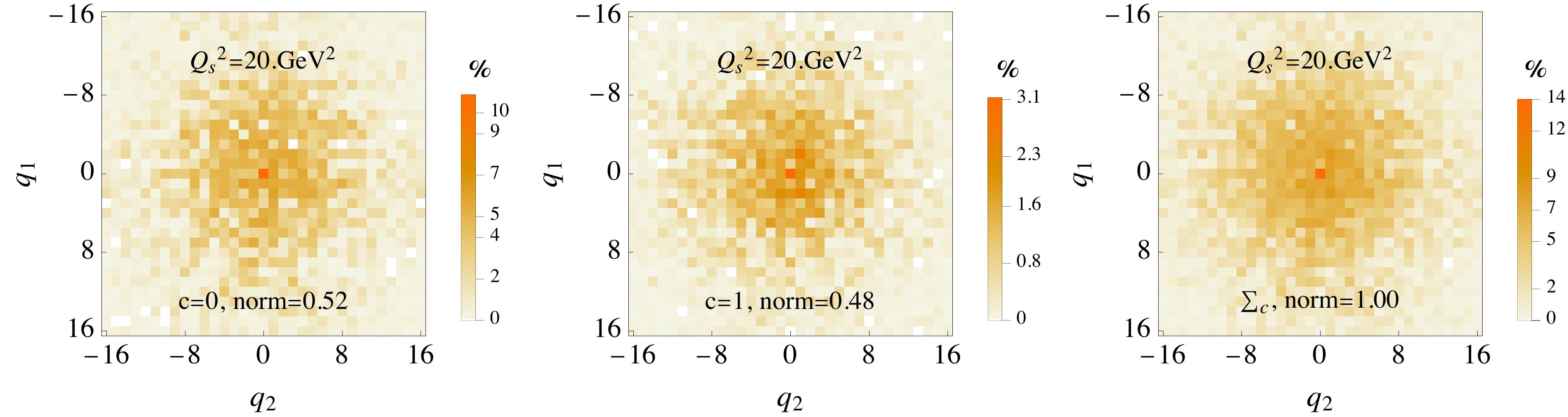}
 }
    \caption{
       Final state probability distribution for $Q_s^2=5.05 \GeV^2$ (a) and $Q_s^2=20.00 \GeV^2$ (b) after evolution in a SU(2) background at infinite jet energy. The first column shows the distribution for the color state $\ket{c}=\ket{0}$, the second column has the result for the color state $\ket{c}=\ket{1}$, and the rightmost columns shows the color summed distributions. The weight of each color to the net distribution is indicated in each plot by ``norm".
       }
    \label{fig:results_2d_plots_SU2}
\end{figure*}

The framework presented in the main text also allows to extract color differential information out of the final quantum state. The capability of accessing the color structure is very important since many jet quenching observables are driven by color flow modifications. In Fig.~\ref{fig:results_2d_plots_SU2}, we show the color differential momentum distributions in transverse space. Since the initial state is a color singlet, the distributions in both color spaces match qualitatively.

\bibliographystyle{apsrev4-1}
\bibliography{Lib.bib}

\end{document}